\documentclass[12pt]{article}
\usepackage{amssymb,amsmath}
\usepackage{amsmath}
\usepackage{amssymb}
\usepackage{graphicx}
\usepackage{tikz}
\usepackage[labelfont=bf]{caption}
\usepackage{color}
\numberwithin{equation}{section}
\newtheorem{theorem}{Theorem}[section]     
\newtheorem{definition}[theorem]{Definition}

\newtheorem{proposition}[theorem]{Proposition}
\newtheorem{lemma}[theorem]{Lemma}

\newtheorem{conjecture}[theorem]{Conjecture}

\newtheorem{remark}[theorem]{Remark}

\def\d{\partial}

\def\p{\partial}
\def\n{\noindent}
\def\f{\frac}

\def\proof{\noindent\hspace{2em}{\itshape Proof: }}
\def\QEDclosed{\mbox{\rule[0pt]{1.3ex}{1.3ex}}} 

\def\QED{\QEDclosed} 
\def\endproof{\hspace*{\fill}~\QED\par\endtrivlist\unskip}
\newcommand{\eqa}{\begin{eqnarray}}
\newcommand{\eeqa}{\end{eqnarray}}
\newcommand{\beq}{\begin{equation}}
\newcommand{\eeq}{\end{equation}}

\newcommand{\ovl}[1]{\ensuremath{\overline{#1}}}
\begin{document}

\title{Integrable viscous conservation laws}
\author{Alessandro Arsie$\;^{a)}$, Paolo Lorenzoni$\;^{b)}$, Antonio Moro$\;^{b,c)}$\\
\\
{\small $\;^{a)}$Department of Mathematics and Statistics}\\
{\small University of Toledo,}
{\small 2801 W. Bancroft St., 43606 Toledo, OH, USA}\\
{\small $\;^{b)}$Dipartimento di Matematica e Applicazioni}\\
{\small Universit\`a di Milano-Bicocca,}
{\small Via Roberto Cozzi 53, I-20125 Milano, Italy} \\
{\small $\;^{c)}$Department of Mathematics and Information Sciences}\\
{\small Northumbria University Newcastle,}
{\small Pandon Building, NE2 1XE, Newcastle upon Tyne, UK}
}

\date{}

\maketitle

\begin{abstract}
We propose an extension of the Dubrovin-Zhang perturbative  approach to the study of normal forms for non-Hamiltonian integrable scalar conservation laws. The explicit computation of the first few corrections leads to the conjecture that such normal forms are parameterized by one single functional parameter, named viscous central invariant.
A constant valued viscous central invariant corresponds to the well-known Burgers hierarchy. The case of a linear viscous central invariant provides a viscous analog of the Camassa-Holm equation, that formerly appeared as a reduction of a two-component Hamiltonian integrable systems. We write explicitly the negative  and positive hierarchy associated with this equation and prove the integrability showing that they can be mapped respectively into the heat hierarchy and its negative counterpart, named the Klein-Gordon hierarchy. A local well-posedness theorem for periodic initial data is also proven. 

We show how transport equations can be used to effectively construct asymptotic solutions via an extension of the quasi-Miura map that preserves  the initial datum.  The method  is alternative to the method of the string equation for Hamiltonian conservation laws and naturally extends to the viscous case.
Using these tools we derive the viscous analog of the Painlev\'e I2 equation that describes the universal behaviour of
the solution at the critical point of gradient catastrophe.
\end{abstract}

\section{Introduction}
In \cite{D} B. Dubrovin proposed a perturbative approach to the study of integrable Hamiltonian evolutionary equations (or systems of equations) of the form
\beq\label{eveq}
u_t=A(u)u_x+\epsilon[B_1(u)u_{xx}+B_2(u)u_x^2]+\epsilon^2[C_1(u)u_{xxx}+C_2(u)u_xu_{xx}+C_3(u)u_x^3]+\mathcal{O}(\epsilon^3)
\eeq
where $\epsilon$ is a small formal expansion parameter. If the unperturbed equation  obtained for  $\epsilon = 0$ is integrable and Hamiltonian (this is always true in the scalar case),  Dubrovin's method provides an effective tool to construct  perturbative dispersive equations of the form~(\ref{eveq}) preserving integrability and Hamiltonian property.
The r.h.s. of \eqref{eveq} is a formal expansion, it contains in principle infinitely many terms and no  additional assumptions concerning convergence are usually enforced. It is natural to classify equations of the form~(\ref{eveq}) up to the group of Miura transformations defined via formal series of the form 
\beq\label{miuratr}
u\to v=f(u)+\epsilon[g(u)u_x]+\epsilon^2[h_1(u)u_{xx}+h_2(u)u_x^2]+...
\eeq
Since most of physical applications  involving equations of the form~(\ref{eveq}) contain a finite number of terms,  one would be tempted 
 to work with truncated Miura transformations only. This choice turns out to be ineffective and the reason is two-fold. First of all, an infinite formal expansion is needed, in general, to define the inverse of a Miura transformation, truncated or not: if one would not allow infinite formal expansion, then the set of truncated Miura transformations would not form a group. Secondly, the restriction to finite expansions will exclude from the analysis  all those non-evolutionary equations that can be put in the evolutionary form \eqref{eveq}.
Let us consider for example the celebrated Camassa-Holm equation
\beq\label{CH}
u_t-\epsilon^2u_{xxt}=-3uu_x+\epsilon^2(uu_{xxx}+2u_xu_{xx}).
\eeq
Introducing the transformation
$$v=u-\epsilon^2u_{xx}$$
its formal inversion is given by an infinite formal series that allows to re-write the equation~(\ref{CH}) in the evolutionary form \eqref{eveq} with respect to the variable $v$.

However in a number of cases,  truncated equations  might have their own interest and in particular those equations possessing  infinitely many
 approximate symmetries \cite{L,S}.
\newline
\newline
In this paper we will extend Dubrovin's approach to the case of integrable conservation laws of the form
\beq\label{eveq2}
u_t=\d_x\left\{u^2+\epsilon[a(u)u_{x}]+\epsilon^2[b_1(u)u_{xx}+b_2(u)u_x2]+\mathcal{O}(\epsilon^3)\right\}.
\eeq
Equations of this form are not necessarily Hamiltonian. For instance, in the scalar case, the operator $\d_x$ defines a Poisson structure on the space of functionals 
$F[u]=\int f(u,u_x,u_{xx},\dots)\,dx$, but in general the current 
$$u^2+\epsilon[a(u)u_{x}]+\epsilon^2[b_1(u)u_{xx}+b_2(u)u_x2]+\mathcal{O}(\epsilon^3)$$
is not the variational derivative of a functional $F[u]$. To overcome this difficulty  we consider the extension of the Hamiltonian formalism to 1-forms introduced in \cite{AL}. Within this framework, currents are viewed as 1-forms in a suitable jet space on which the operator $\d_x$ defines a Poisson structure. Hence, the extension of Dubrovin perturbative scheme is immediate apart from the choice of the class of admissible formal Miura transformations. There are two natural options:
\begin{enumerate}
\item To consider canonical transformations as in~\cite{D} only . This choice is motivated by results in \cite{G} that guarantees any deformation of the operator $\d_x$ to be eliminated by a Miura transformation. 
\item To consider Miura transformations that  preserve the form of the equation \eqref{eveq2}. 
 This  class is clearly wider  and plays a crucial role in the elimination of few inessential functional parameters. This is the reason why we have chosen to work with this second option.
\end{enumerate}
An alternative method based on the existence of formal symmetries of perturbed nonlinear scalar hyperbolic equations has been introduced in~\cite{BGI} (see also~\cite{S}). The classification of integrable hierarchies,  known as {\it symmetry approach}, based instead on the study of nonlinear perturbations to linear equations has been introduced and developed in \cite{MSS,MN,MSY}.

\bigskip

The main results of this paper can be summarized as follows:
\begin{itemize}
\item We classify up to the fifth order in the deformation parameter scalar integrable conservation laws of the form~(\ref{eveq2}).  We will focus on the viscous case  corresponding to $a(u)\ne 0$ and for which the corresponding currents
 are not exact 1-forms.
It turns out that, up to the class Miura transformations preserving the form of the equation \eqref{eveq2}, all deformations are uniquely determined
 by the function $a(u)$ named viscous central invariant. For this reason, we refer to the corresponding equations as \emph{integrable viscous conservation laws}. Dispersive deformations obtained for
$a(u)=0$ and $b_1(u)\ne 0$ have been already studied in \cite{D} in the case of Hamiltonian conservation laws .
\item We observe that except for the case of constant viscous central invariant where the right hand side of \eqref{eveq2} becomes finite and gives the well known  Burgers equation, the generic integrable conservation law contains infinitely many terms.
We show that the case of a linear viscous central invariant $a(u)=u$ corresponds to the equation
\beq\label{newIntro}
u_t-\epsilon u_{xt}=\partial_{x} \left (u^{2} - \epsilon u u_{x} \right),
\eeq
formerly derived as a non-Hamiltonian reduction of two-component generalizations of the Camassa-Holm equation \cite{O96,Fa}. A few interesting properties of equation~(\ref{newIntro}) have been already investigated, such as the existence of a typical viscous shock dynamics associated with cliffons, i.e. weak traveling wave, solutions \cite{Fa}. Nevertheless, its integrable structure has not been yet studied in detail.
We observe that, remarkably, equation~(\ref{newIntro}) is related by a Miura transformation to the equation
\beq\label{chBII}
u_t=\d_x\left(\sum_k \epsilon^k uu_{(k)}\right)
\eeq
which is a symmetry of the equation
 \beq\label{calogero}
u_{\tau_{-1}}  = \partial_{x} \left(\frac{1}{u} + \epsilon  \partial_{x} \frac{1}{u} \right) 
\eeq
previously appeared in a list by F. Calogero~\cite{CAL}. We show that the equation \eqref{calogero} is the first member of the full integrable ``negative" hiearchy 
\beq\label{calogeroHi}
u_{\tau_{-n}}  =\partial_{x} \left( \frac{1}{u} + \epsilon \partial_{x} \frac{1}{u} \right)^{n} (\textup{1}) \qquad n=1,2,3,\dots
\eeq  
We also construct the positive counterpart which turns out to be non-local, that is its evolutionary form is not truncated, as for the equation~(\ref{chBII}).
It should be noted that the negative hierarchy~(\ref{calogeroHi}) and its positive counterparts can be obtained as a one component reduction on the negative hierarchy recursively constructed in~\cite{Fa}. 
We provide an alternative proof of the integrability by showing that the negative and positive hierarchies can be transformed respectively into the Burgers and then the heat hierarchy and its negative counterpart. 
Interestingly, the first member of the negative hierarchy is given by the Klein-Gordon equation and for this reason we refer to it as the Klein-Gordon hierarchy.

\item We study the well-posedness (existence, uniqueness and continuous dependence on initial data) for the periodic Cauchy problem for the equation \eqref{newIntro} (with $\epsilon=1$) locally in time, using the general techniques developed by Kato in \cite{Kato}, similarly to what has been done for the Camassa-Holm equation in \cite{Constantin}. More specifically, we prove that the periodic Cauchy problem for the equation \eqref{newIntro} is locally well-posed for $m\in H^1$, where $m=u-u_x$. It is a work in progress the determination of the conditions that guarantee the global well-posedness of the periodic Cauchy problem and the blow-up conditions in finite time, as it was done for the Camassa-Holm equation~\cite{Constantin, CH}. We note that in \cite{Es} well-posedness and blow-up properties have been studied for the two component generalization of the equation~(\ref{newIntro}) introduced in \cite{Ch} and \cite{Fa}. 

\item Using  {\tt Mathematica} \cite{W} we investigate numerically the Cauchy problem for the equation\eqref{newIntro} assuming periodic boundary conditions. This has been done by converting the equation~(\ref{newIntro}) into a system consisting of an evolutionary PDE together with an ODE. The numerical integrator, based on the method of lines for solving the PDEs, applies a pseudo-spectral method or a spatial discretization with finite differences of the $4$th order. 
We perform numerical simulations with very simple initial data such as translated sinusoidal waves with different wave lengths and observe that the solutions behaviour is as expected: the initial datum evolves in low viscosity regime approaching the gradient catastrophe and the steep profile so created is asymptotically dissipated. 
We also provide numerical evidence of the existence of initial conditions for which the solution breaks in finite time similarly to what happens for the Camassa-Holm equation~\cite{Constantin}. 

\item We give a proof of the quasi-triviality for scalar evolution equations, which is more elementary and alternative to the one provided in \cite{LZ}. 
Inspired by the work \cite{BGI}, our approach is based on the semiclassical expansion of solutions. We show how to use transport equations to construct an extension of the quasi-Miura transformation that preserves the initial datum. 
In the case of the Burgers equation, we find recurrence relations at all orders for the aforementioned transformation, and provide evidence that the obtained asymptotic expansion coincides with the one constructed via the Laplace method, starting from the solutions of the heat hierarchy. 

\item The class of perturbed equations considered in this paper coincides up to $O(\epsilon^{2})$ with the generalized Burgers equations studied in \cite{DE}. We show that, near the critical point, the solutions of scalar non Hamiltonian conservation laws should  satisfy a second order ODE, 
 which is the non-Hamiltonian analogue of the Painlev\'e equation arising in the description
 of the critical behaviour of Hamiltonian conservation laws \cite{D}. This ODE admits the Pearcey function as a particular solution. This is consistent with the results in~\cite{I}  and \cite{DE} on the universality of the critical behaviour of the (generalized) Burgers equation.
\end{itemize}

The paper is organized as follows. In Section 2 we give the definition of integrable scalar conservation laws. We introduce their normal form in Section 3. In Section 4 we present perturbative classification results concerning the integrable viscous conservation laws. A detailed discussion of the case of the linear viscous central invariant in provided in Section 5. Section 6 is devoted to the proof of the local well-posedness for the Cauchy problem and numerical simulations are presented in Section 7. In Section 8 we give an elementary proof of the quasi-triviality of scalar evolutionary PDEs. 
Instead of characterizing the quasi-triviality transformations in terms of their infinitesimal generators as in \cite{LZ}, we provide a direct characterization of the coefficients of the transformation in terms of solutions of transport equations, in the spirit of the work~\cite{BGI}. In Section 9, above mentioned results are applied  to the study of the critical behaviour of solutions near the critical point. 

\section{Commuting flows}
In the present section, we introduce the main notions concerning integrability and perturbation theory of conservation laws. For the sake of simplicity we specify all definitions and properties in the case of scalar conservation laws, that is the subject matter of this paper.
 The interested reader will find a more general discussion in the paper \cite{AL}.

Let
\beq
\label{cons1}
u_{t} =\partial_{x} \alpha \left(u,u_{x},u_{xx}, \dots \right)
\eeq
be a scalar conservation law. If we assume that the current $\alpha$ is a differential polynomial, then, rescaling the variables as $x\to\epsilon x,t\to\epsilon t$, we can write the right hand side of any scalar conservation law as
\beq
\label{cons1}
u_{t} =\partial_{x}\left(\alpha_0(u)+\sum_{k=1}^N\epsilon^k\alpha_k(u,u_x,u_{xx},...) \right)
\eeq
where $\alpha_k$ are differential polynomyals of degree $k$. Here the degree is assigned according
 to the following rule: ${\rm deg}\,f(u)=0,\,{\rm deg}\,u_{(k)}=k$. Clearly if we start from a differential polynomial, $N$ is finite. However as mentioned above we also include 
  the case $N=\infty$. In this case the equation \eqref{cons1} will be referred to as \emph{formal
   conservation law}.

\begin{definition}
A (formal) scalar conservation law \eqref{cons1}  is said to be integrable if it admits infinitely many symmetries of the form
\beq
\label{cons2}
u_{\tau} =\partial_{x} \beta \left(u,u_{x},u_{xx}, \dots \right).
\eeq
\end{definition}

A very important subcase is given
 by the class of Hamiltonian conservation laws. In this case the current $\alpha$ is  the variational
  derivative of  a suitable local functional (the Hamiltonian functional)
 $$H[u]=\int h(u,u_x,...)\,dx$$
 where $h(u,u_x,...)$ is the Hamiltonian density. In this case we have
 $$\alpha=\f{\delta H}{\delta u}=\f{\d h}{\d u}-\d_x\left(\f{\d h}{\d u_x}\right)+\d^2_x\left(\f{\d h}{\d u_{xx}}\right)+\dots.$$
 Two functionals $F[u]$ and $G[u]$ such that $\{F,G\}=0$ are said to be \emph{in involution}.
 
 It is well-known (see e.g.\cite{DN}) that the operator $\d_x$ defines a Poisson bracket in the space
  of local functionals. Given two local functionals $F[u]$ and $G[u]$, their Poisson bracket
   is the local functional defined by
\beq\label{DNbracket}   
\{F[u],G[u]\}=\int \f{\delta F}{\delta u}\d_x\left(\f{\delta G}{\delta u}\right)\,dx.
\eeq
It was proved in~\cite{G,DMS} that any local Hamiltonian operator can be reduced in the scalar case to $\d_x$ by means of a Miura transformation. This means that any scalar Hamiltonian equation (w.r.t a local Hamiltonian operator) can be written as a conservation law after a suitable Miura trasformation. 
  
 The previous analysis shows that the function $\alpha$ in
   \eqref{cons1} can be thought as a 1-form. Such a 1-form is in general non-exact and the equation is not Hamiltonian with respect to the Hamiltonian operator $\d_x$. 
  
 Nevertheless, extending a construction that is well-known in the finite dimensional setting, it is possible to define a Poisson bracket 
 on the space of 1-forms $\Lambda_1$. In the scalar case, this is given by
\beq
\label{ALbracket}
\{\alpha,\beta \} := \sum_{j} \partial_{x}^{j+1} \beta \frac{\partial \alpha}{\partial u_{(j)}} - \partial_{x}^{j+1} \alpha \frac{ \partial \beta}{\partial u_{(j)}}  =0.
\eeq    
The above bracket satisfies the following properties (see \cite{AL} for more details):
\begin{enumerate}
\item If $\alpha=\delta F$ and $\beta=\delta G$, then $\{\alpha, \beta\}=\delta \{F,G\}.$
\item $\{\cdot,\cdot\}$ equips the space of $1$-forms $\Lambda_1$ with a Lie algebra structure;
\item the Poisson structure induces an (anti)-homomorphism of Lie algebras between $(\Lambda_1, \, \{\cdot, \cdot\})$ and the space of evolutionary vector fields equipped with the Lie bracket given by the Lie commutator. 
\end{enumerate}
From the last property it follows that two scalar conservation laws commute if and only if the associated currents are in involution.
For convenience of the reader we give an alternative elementary proof of this fact.  
\begin{proposition}
Two flows of conservation laws of the form~\eqref{cons1} and~\eqref{cons2} commute iff $\alpha$ and $\beta$ are in involution with respect to the bracket \eqref{ALbracket}.
\end{proposition}
\proof
Let us introduce the function $\varphi$ such that $u = \varphi_{x}$. Then, integrating once w.r.t. the variable $x$, equations~(\ref{cons1}) and~(\ref{cons2}) give
\begin{align*}
\varphi_{t} =& \alpha(u,u_{x},u_{xx},\dots) + f(t,\tau) \\
\varphi_{\tau} =& \beta(u,u_{x},u_{xx},\dots)+ g(t,\tau)
\end{align*}
where $f$ and $g$ are two arbitrary functions. The request that the two equations above commute for any solution $\varphi$, i.e.
\[
\partial_{\tau} \varphi_{t} = \partial_{t} \varphi_{\tau}
\]
implies that the two compatibility conditions 
\begin{equation*}
\alpha_{\tau} = \beta_{t} \qquad f_{\tau} = g_{t}
\end{equation*}
must hold separately. The second condition is identically satisfied if one choses $f =\psi_{t}$ and $g = \psi_{\tau}$. Using the chain rule together with the equations~(\ref{cons1}) and~(\ref{cons2}) the first condition can be equivalently written as follows
\begin{equation*}
\alpha_{\tau} - \beta_{t} = \sum_{j} \partial_{x}^{j+1} \beta \frac{\partial \alpha}{\partial u_{(j)}} - \partial_{x}^{j+1} \alpha \frac{ \partial \beta}{\partial u_{(j)}}  =0.
\end{equation*}
The proposition is proved.
\endproof

As a consequence of the above proposition, we have that any integrable hierarchy of conservation laws is always defined by an infinite family of 1-forms in involution. From this perspective Hamiltonian and non-Hamiltonian conservation laws can be understood within the same framework.

{\bf Example} Let us consider the Burgers hierarchy defined as 
\beq
u_{t_n}=\d_x \omega_n=\d_x\left[\left(u + \d_x \right)^n u\right],\qquad n=0,1,2,...
\eeq
The flows associated with $\omega_{n}$ are in involution w.r.t. the Poisson bracket \eqref{ALbracket}, i.e. $\{\omega_n, \omega_m\}=0$.

\section{The normal form of conservation laws}

Let us consider a formal conservation law of the form
\beq
\label{pert}
u_t= \partial_{x} \left(u^{2} + \epsilon a(u) u_{x} + \epsilon^{2} (b_{1}(u) u_{xx} + b_{2}(u) u_{x}^{2}) + O(\epsilon^{3}) \right).
\eeq
It can be viewed as a higher order perturbation of the Hopf equation 
\beq
\label{hopf}
u_t=\d_x u^2 =2uu_x.
\eeq
Let us observe that there is no loss of generality by taking the Hopf equation as the leading order. Indeed, any equation of the form $u_{t} =K(u) u_{x}$ can be transformed into the equation~(\ref{hopf}) by a re-parametrization of the dependent variable $u = u(v)$.
The Hopf equation~(\ref{hopf}) is Hamiltonian in the standard sense and completely integrable. Hence, there exist an infinite set of symmetries parametrized by a function of one variable $f(u)$ of the form
\beq\label{sym}
u_{\tau}=\partial_{x} f(u) = f'(u)u_x.
\eeq
It is straightforward to check that $\partial_{t} u_{\tau} = \partial_{\tau} u_{t}$ for any function $f(u)$.

The aim of this section is to illustrate a perturbative approach to the problem of classifying
 integrable deformations of the Hopf hierarchy of the form~(\ref{pert}).
Various approaches have been developed so far to construct such deformations (see for instance  \cite{BGI,DZ,L,S,D,LZ}).

%


The approach we propose here exploits the conservative form of the Hopf hierarchy and it
 is based on the additional requirement that also its deformations maintain this property. In other words
 we  deform the currents viewed as 1-forms, without requiring them to be exact.

The classification procedure discussed in this paper is based on the following:

\begin{definition}
\label{approxinteg}
The conservation law associated to the 1-form 
\[
\omega^{def}_{f} = f(u) + \sum_{n=1}^{\infty} \epsilon^{n} g_{n}(u, u_{x},\dots)
\]
where each $g_{j}$ is a homogenous differential polynomial of order $j$, is said to be integrable up to the order $\epsilon^{k}$ if there exists a 1-form 
\[
\omega^{def}_{\tilde{f}} = \tilde{f}(u) + \sum_{n=1}^{\infty} \epsilon^{n} \tilde{g}_{n}(u, u_{x},\dots)
\]
such that their Poisson bracket  $\{\omega^{def}_f,\omega^{def}_{\tilde{f}}\}=0 $ vanishes identically modulo $O(\epsilon^{k+1})$ for any function $\tilde{f}(u)$.
\end{definition}
Let us observe that taking for instance $f(u) = u^{2}$, the request that the two forms $\omega_{u^{2}}^{def}$ and $\omega_{\tilde{f}}^{def}$ are in involution for any ${\tilde f}(u)$ up to a certain order in $\epsilon$ means that the corresponding deformed conservation law inherits all the symmetries of the Hopf equation up to the same order.

In the case of exact 1-forms the above involutivity condition reduces to the standard involutivity condition between Hamiltonian functionals with respect to the canonical Poisson bracket (see e.g.~\cite{D}). 

A direct application of the definition~(\ref{approxinteg}) provides an effective tool to classify approximate integrable conservation laws viewed as deformations to the Hopf equation. The explicit derivation of integrability conditions becomes computationally more and more expensive as the order of such deformations increases. Nevertheless, it turns out that the most general deformation of the Hopf equation contains a certain number of redundant functional parameters, that can be eliminated using the invariance of the form of conservation laws with respect to a special class of Miura transformations.

For instance, let us consider a general deformed conservation law of the form
\beq
\label{hopfdef}
u_t=\d_x\omega^{def}_{u^2}(u,u_x,\dots;\epsilon)
\eeq
with 
\begin{gather*}
\begin{aligned}
\omega^{def}_{u^2} &= u^{2} + \epsilon a(u) u_{x} + \epsilon^{2} \left(b_{1}(u) u_{xx}  + b_{2}(u) u_{x}^{2} \right) + \epsilon^{3} \left( c_{1}(u) u_{xxx} + c_{2}(u) u_{x} u_{xx} + c_{3}(u) u_{x}^{3} \right) \\
&+ \epsilon^{4} \left (d_{1}(u) u_{4x} + d_{2}(u) u_{x} u_{xxx} + d_{3}(u) u_{xx}^{2} + d_{4}(u) u_{x}^{2} u_{xx} + d_{5}(u) u_{x}^{4} \right) + \epsilon^{5} \left(e_{1}(u) u_{5x}  \right . \\ 
 & \left . + e_{2}(u) u_{x} u_{4x} + e_{3}(u) u_{xx} u_{xxx}  + e_{4}(u) u_{x}^{2} u_{xxx} + e_{5}(u) u_{x} u_{xx}^{2} + e_{6}(u) u_{x}^{3} u_{xx} + e_{7}(u) u_{x}^{5} \right)   \\ 
 &+ O(\epsilon^{6}).
\end{aligned}
\end{gather*}
We observe that the form of the equation~(\ref{hopfdef}) is preserved under the Miura transformation  
\beq
\label{miura}
u\to v=u+\epsilon^k\d_x\beta(u,u_x,\dots)
\eeq
where $\beta$ is a homogeneous differential polynomial of degree $k-1$.  Hence, applying the Miura transformation~(\ref{miura}) to the equation~(\ref{hopfdef}) we get
$$v_t=\d_x\left(\omega_{v^2}(u(v,v_x,...),u_x(v,v_x,...),...)+\beta_t(u(v,v_x,...),u_x(v,v_x,...),...)\right),$$
where
$$u(v,v_x,...)=v-\epsilon^k\d_x\beta(v,v_x,\dots)+O(\epsilon^{k+1}).$$

Clearly the transformation does not affect the terms of the equation of order less then $k$
 in $\epsilon$. At the order $k$ we have
$$\omega_k\to\tilde{\omega}_k=\omega_k(v,v_x,...)-2v\d_x\beta(v,v_x,...)
+\sum_{s=0}^{k-1}\f{\d\beta(v,v_x,...)}{\d v_{(s)}}\d_x^{s+1}v^2$$
or equivalently
\begin{equation}\label{miuraform}
\tilde{\omega}_k=\omega_k(v,v_x,...)+\sum_{s=1}^{k-1}\f{\d\beta(v,v_x,...)}{\d v_{(s)}}\left[\sum_{l=1}^s\binom{s+1}{l}\,v_{(l)}v_{(s+1-l)}\right].
\end{equation}
For instance, for $k=2$ it is $\beta=\beta_1v_x$ and  
$$\tilde{\omega}_2=b_1v_{xx}+(b_2+2\beta_1)v_x^2.$$
Hence the term $v_{x}^{2}$ can be eliminated by choosing $\beta_{1} = - b_{2}/2$.
For $k=3$ we have  $\beta=\beta_{21}v_{xx}+\beta_{22}v_x^2$ and
$$\tilde{\omega}_3=c_{1}(v) v_{xxx} + (c_{2}+6\beta_{21})v_{x}v_{xx} + (c_{3}+4\beta_{22})v_x^3.$$
Analogously, both terms $v_{x} v_{xx}$ and $v_{x}^{3}$ can be eliminated.
Similarly, it is straightforward to check that for $k=4$ one can retain just the terms $v_{4x}$ and $v_{x} v_{xx}$, or alternatively $v_{xx}^{2}$, while for $k=5$ only $v_{5x}$ and $v_{x} v_{4x}$, or alternatively $v_{xx} v_{xxx}$, will survive after a suitable Miura transformation.

The above examples suggest that it should be possible to choose $\beta$ in such a way that only those terms that do not factor through $v_{x}$ survive after the action of a suitable Miura transformation. This turns out to be true:

\begin{theorem}\label{THnormalform} 
Given the conservation law 
\begin{equation}\label{clold} 
u_t=\d_x\omega_{u^{2}} 
\end{equation} 
with $\omega_{u^{2}} = u^{2} + \epsilon a(u)u_x+\sum_{k \geq2} \epsilon^{k} \omega_k(u,u_{x}, \dots)$ there exists a sequence of Miura transformations of the form 
\[ 
u  \to v=u+ \epsilon^k \partial_{x}\beta (u,u_x,\dots) 
\] 
that brings equation \eqref{clold} into the normal form
\begin{equation}\label{clnew} 
v_t=\d_x \omega_{v^{2}} 
\end{equation} 
where 
$$\omega_{v^{2}} = v^{2} + \epsilon a(v) v_x+\sum_{k>1} \epsilon^{k} \omega_k(v, v_{x}, \dots)
$$ 
is such that 
$$
\frac{\partial \omega_k}{\d v_{x}}=0, \quad \forall k>1.
$$ 
\end{theorem} \proof
Let us first introduce some useful notations. A general 1-form $\beta$ homogeneous of degree $k-1$ will be denoted as follows:
$$\sum_{i_1, \dots, i_{k-1}\; \;
i_1+2i_2+\dots+(k-1)i_{k-1}=k-1}\beta_{[i_1, i_2, \dots, i_{k-1}]}v_{(1)}^{i_1}\dots v_{(k-1)}^{i_{k-1}}.$$ 
We will prove that the coefficients $\beta_{[i_1, i_2, \dots, i_{k-1}]}$ can be are recursively determined  through a lower triangular relation with respect to a suitable ordering of the monomials $v_{(1)}^{i_1}\dots v_{(k-1)}^{i_{k-1}}$. 
We introduce such an ordering, which is similar to a reverse lexicografic ordering: the monomial $v_{(1)}^{i_1}\dots v_{(k-1)}^{i_{k-1}}$ ranks higher (or comes first or has a higher rank) than $v_{(1)}^{j_1}\dots v_{(k-1)}^{j_{k-1}}$, symbolically
$$
v_{(1)}^{i_1}\dots v_{(k-1)}^{i_{k-1}}\succ v_{(1)}^{j_1}\dots v_{(k-1)}^{j_{k-1}},
$$ 
if  there exists $m\in \{1, \dots, k-1\}$ such that $i_{l}=j_{l}$ for all $l>m$ and $i_m>j_m$. 
In other words, monomials are ranked according to the higher derivative of $v$. For instance, among the homogenous monomials of degree $k-1$, the highest ranking monomial is $v_{(k-1)}$ and the lowest ranking is $v_{(1)}^{k-1}$.  This gives clearly a total ordering of the monomials. 
In the following, we will write down a 1-form homogeneous of a certain degree from the highest to the lowest ranking. 
For instance, for a homogeneous 1-form of degree $3$ 
$$\beta=\beta_{[0,0,1]}v_{(3)}+\beta_{[1,1,0]}v_{(1)}v_{(2)}+\beta_{[3,0,0]}v_{(1)}^3.$$

Let us now fix $k$ and write $\omega_k$ as $\alpha +v_{(1)}\eta$ where $\eta$ is ordered as above. In the rest of the proof $\alpha$ will be neglected as we will focus only on the terms of $v_{(1)}\eta$ that factors through $v_{(1)}$.

Let us now expand the expression 
\begin{equation}\label{miuraform2}\sum_{s=1}^{k-1}\f{\d\beta(v,v_x,...)}{\d v_{(s)}}\left[\sum_{l=1}^s\binom{s+1}{l}\,v_{(l)}v_{(s+1-l)}\right]
\end{equation}
entering the Miura transformation \eqref{miuraform} as follows
$$
\sum_{s=1}^{k-1}\f{\d\beta(v,v_x,...)}{\d v_{(s)}}2(s+1)v_{(s)}v_{(1)}+\sum_{s=3}^{k-1}\f{\d\beta(v,v_x,...)}{\d v_{(s)}}\left[\sum_{l=2}^{s-1}\binom{s+1}{l}\,v_{(l)}v_{(s+1-l)}\right]
$$
$$
=\tilde\beta v_{(1)}+\delta,
$$
where 
$$\tilde\beta=\sum_{s=1}^{k-1}\f{\d\beta(v,v_x,...)}{\d v_{(s)}}2(s+1)v_{(s)}
$$ is meant to be ordered as above and the residual term is
\beq
\label{delta}
\delta =\sum_{s=3}^{k-1}\f{\d\beta(v,v_x,...)}{\d v_{(s)}}\left[\sum_{l=2}^{s-1}\binom{s+1}{l}\,v_{(l)}v_{(s+1-l)}\right].
\eeq
Notice that the  coefficients in $\tilde \beta$ coincide, up to a positive factor, with the coefficients in $\beta$, that is
$$
\tilde\beta_{[i_1, i_2, \dots, i_{k-1}]}=c_{i_1, \dots, i_{k-1}}\beta_{[i_1, \dots, i_{k-1}]},
$$
where $c_{i_1, \dots, i_{k-1}}$ are positive constants. In particular, terms in $\tilde\beta$ have the same ranking of the corresponding terms in $\beta$.

We are now able to prove that it is always possible to eliminate all terms containing $v_{(1)}$ starting from the highest ranking coefficients. However, special attention should be paid as one needs to take into account contributions coming from both $\tilde \beta v_{(1)}$ and $\eta v_{(1)}$ and $\delta$. 
It will turn out that all such contributions appear in a lower triangular form and then we can recursively fix coefficients of $\tilde\beta$ and consequently $\beta$ in such a way to kill  $\eta v_{(1)}$.

Let us first note that is is always possible to fix the highest ranking coefficient in $\tilde \beta v_{(1)}$ that annihilates the highest ranking coefficient in $v_{(1)}\eta$. Indeed, the highest ranking term in $\tilde \beta v_{(1)}$ is $\tilde \beta_{[0, 0, \dots, 1]}v_{(1)}v_{(k-1)}$ while the highest ranking term in $\eta v_{(1)}$ is $\eta_{[0, 0, \dots, 1]} v_{(1)}v_{(k-1)}$. It is immediate to check that no such a term can originate from $\delta$ and therefore $\tilde \beta_{[0, 0, \dots, 1]}$ is uniquely determined by the condition $\tilde \beta_{[0, 0, \dots, 1]}+\eta_{[0, 0, \dots, 1]}=0$. 

Let us now proceed by induction following the ordering (from the highest to the lowest ranking) of the monomials. Assuming that the $L$ highest ranking monomial in $\tilde \beta$ is determined, we show that it is possible to find the $L+1$ highest ranking monomial  $\tilde \beta_{[i_1, \dots, i_{k-1}]}v_{(1)}^{i_1+1}\dots v_{(k-1)}^{i_{k-1}}$ in $\tilde \beta$ that annihilates the corresponding $L+1$ highest ranking monomial $\eta_{[i_1, \dots, i_{k-1}]}v_{(1)}^{i_1+1}\dots v_{(k-1)}^{i_{k-1}}$ in $\eta v_{(1)}$ by imposing the condition 
\begin{equation}
\label{eqconresidual}\eta_{[i_1, \dots, i_{k-1}]}v_{(1)}^{i_1+1}\dots v_{(k-1)}^{i_{k-1}}+\tilde \beta_{[i_1, \dots, i_{k-1}]}v_{(1)}^{i_1+1}\dots v_{(k-1)}^{i_{k-1}}+\text{ residual terms} =0,
\end{equation}
where residual terms come from $\delta$. Indices $i_1, \dots, i_{k-1}$ above are fixed.  

From its definition, $\delta$ can be thought as an operator acting linearly on $\beta$, that is $\delta(\beta_1+\beta_2) = \delta(\beta_1)+\delta(\beta_2)$. Therefore, without loss of generality, we can focus our attention on the action of $\delta$ on the generic term of the form
 $$\gamma=\beta_{[j_1, \dots, j_{k-1}]}v_{(1)}^{j_1}\dots v_{(k-1)}^{j_{k-1}}.$$ 
 Up to a positive factor, we can write 
 \begin{equation}\label{orderingkey}\delta(\gamma)=\beta_{[j_1, \dots, j_{k-1}]}\sum_{s=3}^{k-1}\sum_{l=2}^{s-1}v_{(1)}^{j_1}\dots v_{(l)}^{j_l+1}\dots v_{(s+1-l)}^{j_{s+1-l}+1}\dots v_{(s)}^{j_s-1}\dots v_{(k-1)}^{j_k-1}.\end{equation}
Since $l\leq s-1<s$ and analogously $s+1-l < s$ and  $j_s$ is mapped into $j_s-1$, from equation \eqref{orderingkey} the operator $\delta$ always decreases the ranking of the terms on which it acts.
 
This means that the residual term into the equation \eqref{eqconresidual} only contains coefficients in $\beta$ (i.e $\tilde \beta$) already determined when considering the corresponding equation at the higher orders.
Hence we can solve the linear equation\eqref{eqconresidual} w.r.t  $\beta_{[i_1, \dots, i_{k-1}]}$ in terms of $\eta_{[i_1, \dots, i_{k-1}]}$ and in terms of  already determined higher ranking coefficients. This concludes our proof.

\endproof

\section{Classification results}
\label{secclass}
We now proceed to the classification of involutive 1-forms
\begin{gather}
\label{normalform}
\begin{aligned}
\omega^{def}_{u^2} &= u^{2} + \epsilon a(u) u_{x} + \epsilon^{2}b_{1}(u) u_{xx} +\epsilon^{3}c_{1}(u) u_{xxx}+ \epsilon^{4} 
\left[d_{1}(u) u_{4x} + d_{2}(u) u_{xx}^2\right]+\\ 
&+ \epsilon^{5} \left[e_{1}(u) u_{5x} + e_{2}(u) u_{xx} u_{xxx}\right]+\dots \\
\end{aligned}
\end{gather}
\begin{gather}
\label{normaldeformed}
\begin{aligned}
\omega^{def}_f &=  f(u) + \epsilon Au_{x} + \epsilon^{2} \left(B_{1}u_{xx}  + B_{2}u_{x}^{2} \right) + \epsilon^{3} \left( C_{1}u_{xxx}
+ C_{2}u_{x} u_{xx} + C_{3}u_{x}^{3} \right) \\
&+ \epsilon^{4} \left (D_{1}u_{4x} + D_{2}u_{x} u_{xxx} + D_{3}u_{xx}^{2} + D_{4}u_{x}^{2} u_{xx} + D_{5}u_{x}^{4} \right) \\
&+ \epsilon^{5} \left(E_{1}u_{5x} + E_{2}u_{x} u_{4x} + E_{3}u_{xx} u_{xxx} +E_{4}u_{x}^{2} u_{xxx} 
+ E_{5}u_{x} u_{xx}^{2} + E_{6}u_{x}^{3} u_{xx} + E_{7}u_{x}^{5} \right)\\
&+\dots
 \end{aligned}
\end{gather}
up to a certain order $\epsilon^{k}$ w.r.t. the Poisson bracket defined in~(\ref{ALbracket}), that is  
$$
\{\omega^{def}_{u^2},\omega^{def}_{f} \} =O(\epsilon^{k+1}).
$$
In virtue of the theorem~(\ref{THnormalform}) we can take $\omega_{u^{2}}^{def}$ in its normal form~(\ref{normalform}) without loss of generality. We have the following

\begin{theorem}\label{classth}
If $a(u)\ne 0$, then up to $O(\epsilon^{6})$, the normal forms of integrable conservation laws
\[
u_{t} = \partial_{x} \omega_{u^{2}}^{def}
\]
and their commuting flows
\[
u_{\tau} = \partial_{x} \omega_{f}^{def}
\]
 are parameterized only by the functional parameter $a(u)$ (here $\omega_{u^{2}}^{def}$ and $\omega_{f}^{def}$ are respectively given by~(\ref{normalform}) and~(\ref{normaldeformed}))\footnote{This statement has been proven to hold up to the order $O(\epsilon^{12})$ by J. Ekstrand\cite{JE}}. Moreover, up to $O(\epsilon^{6})$,  two scalar conservation laws sharing one and the same inviscid limit are Miura equivalent if and only if they are associated to the same functional parameter $a(u)$.
\end{theorem}

\noindent
\proof
The constraint that the Poisson bracket $\{\omega^{def}_{u^2},\omega^{def}_{f} \} $ vanishes up to the order $O(\epsilon^{6})$ provides the following set of conditions on the deformation coefficients \\

\noindent $O(\epsilon^{0}) \to $ no conditions  \\

\noindent $O(\epsilon) \to$ Gives $A$ in terms of $a$ and $f$: \\
\begin{equation}
A(u) = \frac{1}{2} a(u) f''(u)
\end{equation}
$O(\epsilon^{2}) \to$ Gives $B_{1}$ and $B_{2}$ in terms of $b_{1}$, $f$ and $a(u)$:

\begin{eqnarray*}
B_1&=&\f{1}{2}b_1f''+\f{1}{6}a^{2}f'''\\
B_2&=&\f{1}{4}aa'f'''+\f{1}{8}a^2f^{(4)}+\f{1}{4}b_1f'''
\end{eqnarray*}

\noindent $O(\epsilon^{3})\to$ Gives $C_{1}$, $C_{2}$ and $C_{3}$ in terms of the small letters and $f$
\begin{eqnarray*}
C_1&=&\f{1}{3}a^2a'f'''+\f{1}{2}c_1f''+\f{1}{24}a^3f^{(4)}\\
C_2&=&\f{11}{12}a(a')^2f'''+\f{5}{6}a^2a'f^{(4)}+\f{7}{24}a^2a''f'''+\f{3}{4}c_1f'''+\f{1}{12}a^3f^{(5)}\\
C_3&=&\f{1}{3}aa'a''f'''+\f{11}{24}a(a')^2f^{(4)}+\f{1}{6}c_1f^{(4)}+\f{1}{48}a^3f^{(6)}+\f{1}{18}a^2a'''f'''
+\f{1}{6}a^2a''f^{(4)}+\f{1}{4}a^2a'f^{(5)} 
\end{eqnarray*}
together with the constraint  \\
\begin{equation}
b_1=(a^2/2!)'
\end{equation}
\noindent $O(\epsilon^{4})\to$ Gives $D_{1}$, $D_{2}$, $D_{3}$, $D_{4}$ and $D_{5}$ in terms of the small letters and $f$ 
\begin{eqnarray*}
D_1&=&\f{1}{8}a^3a'f^{(4)}+\f{1}{6}a^3a''f'''+\f{1}{120}a^{(4)}f^{(5)}+\f{1}{2}a^2(a')^2f'''+\f{1}{2}d_1f''\\
D_2&=&\f{9}{16}a^3a''f^{(4)}+\f{1}{2}d_2af''+\f{7}{4}a^2a'a''f'''+d_1f'''+\f{1}{6}a^3a'''f'''+
\f{1}{48}a^4f^{(6)}+\f{3}{8}a^3a'f^{(5)}+\\
&&+\f{15}{8}a^2(a')^2f^{(4)}+\f{3}{2}a(a')^3f'''\\
D_3&=&\f{17}{24}a^2a'a''f'''+\f{1}{72}a^3a'''f'''+\f{17}{48}a^3a''f^{(4)}+\f{11}{12}a(a')^3f'''+\f{5}{4}a^2(a')^2f^{(4)}
+\f{3}{4}d_1f'''+\\
&&+\f{1}{	72}a^4f^{(6)}+\f{1}{4}a^3a'f^{(5)}\\  
D_4&=&\f{7}{16}a^3a'f^{(6)}+\f{3}{4}a^3a''f^{(5)}+\f{29}{30}a^2a'a'''f'''+\f{27}{8}a(a')^3f^{(4)}+\f{1}{48}a^4f^{(7)}+\f{3}{5}d_2f'''+\\
&&+\f{29}{10}a(a')^2a''f'''+4a^2a'a''f^{(4)}+d_1f^{(4)}+\f{21}{8}a^2(a')^2f^{(5)}+\f{1}{3}a^3a'''f^{(4)}+\\
&&+\f{1}{12}a^3aa^{(4)}f'''
+\f{9}{10}a^2(a'')^2f'''\\
D_5&=&\f{23}{576}a^3a^{(4)}f^{(4)}+\f{1}{144}a^3a^{(5)}f'''+\f{19}{48}a^2(a'')^2f^{(4)}+\f{1}{8}d_2f^{(4)}+\f{1}{8}a^3a''f^{(6)}+\f{13}{144}a^3a'''f^{(5)}
+\\
&&+\f{3}{4}a(a')^3f^{(5)}+\f{1}{384}a^4f^{(8)}+\f{23}{144}a^2a''a'''f'''+\f{7}{16}a^2(a')^2f^{(6)}+\f{3}{16}aa'(a'')^2f'''+\\
&&+\f{1}{16}a^3a'f^{(7)}+
\f{1}{8}d_1f^{(5)}+\f{73}{144}a^2a'a'''f^{(4)}+\f{13}{144}a^2a'a^{(4)}f'''+\f{47}{48}a^2a'a''f^{(5)}+\\
&&+\f{7}{18}a(a')^2a'''f'''+\f{4}{3}a(a')^2a''f^{(4)}.
\end{eqnarray*}
and the constraint
\begin{equation}
c_1=(a^3/3!)''
\end{equation}
\noindent $O(\epsilon^{5})\to$ Gives $E_{1}$, $E_{2}$, $E_{3}$, $E_{4}$, $E_{5}$, $E_{6}$ and $E_{7}$ in terms of the small letters and $f$
 (whose cumbersome expression is not presented here) and the constraints
\begin{eqnarray*}
d_1&=&(a^4/4!)'''\\
d_2&=&\f{5}{24}a^3a^{(4)}+\f{8}{3}a(a')^2a''+a^2(a'')^2+\f{31}{18}a^2a'a'''. 
\end{eqnarray*}
It is immediate to check that coefficient $a(u)$ is invariant w.r.t subgroup of Miura transformations  that preserve the inviscid limit. This proves the last part of the statement.
\endproof
We call {\it integrable viscous conservations laws} all those  conservation laws that are obtained via this deformation procedure extended at any order in the parameter $\epsilon$ and  such that $a(u) \neq 0$.
The perturbative calculations performed above support the formulation of the following
\begin{conjecture}\label{maincon}
The normal form of integrable viscous conservation laws associated with the 1-form~(\ref{normalform}) as well as their commuting flows associated with the 1-form~(\ref{normaldeformed}) are uniquely determined at any order in $\epsilon$ by the non-vanishing functional parameter $a(u)$. 
\end{conjecture}
It should also be noted that if $a(u) $ is allowed to be vanishing the deformation procedure develops a branching. The branch $a(u) =0$ corresponds to the case of dispersive perturbations. In particular, Hamiltonian dispersive perturbations have been extensively studied in a number of papers (see e.g.~\cite{DZ,D,LZ}). It is conjectured that all deformations are uniquely specified by a number of functional parameters that 
 in the bi-Hamiltonian setting are called central invariants \cite{DLZ}. The functional parameter $a(u)$  plays in the present context the same role as central invariants in the Hamiltonian setup. Hence, we refer to $a(u)$ as {\it viscous central invariant}.


An important well-known example of integrable viscous conservation law as been already mentioned above and it is given by the Burgers equation
\beq
\label{burgers}
u_{t} = \partial_{x} \left( u^{2} + \epsilon u_{x} \right) = 2 u u_{x} + \epsilon u_{xx},
\eeq
Let us observe that Burgers' equation already appears in its normal form and it possesses the constant viscous central invariant $a(u) = 1$.
Conversely,  in the case of constant viscous central invariant the 1-form $\omega_{u^{2}}^{def}$ truncates at the first order in $\epsilon$ and gives the Burgers equation. Higher flows of the Burgers hierarchy can be obtained as a deformation of  higher flows of the Hopf hierarchy obtained by choosing in~(\ref{normaldeformed}) $f(u) = u^{n} $, $n \geq 2$. 
\begin{remark}
In the bi-Hamiltonian setting, the case of constant central invariants, that corresponds to the KdV hierarchy, plays a crucial role in two dimensional quantum gravity~\cite{Witten,Kont}. In particular, the generating function of correlators  can be obtained from a solution of the hierarchy specified by a suitable initial datum. 

Similarly, here we point out that the free energy for a generalized Curie-Weiss model in Statistical Mechanics can be viewed as a particular  solution of the full potential Burgers hierarchy. We should also point out that the connection between statistical mean field models and the Burgers equation has been already observed in the literature (see e.g. \cite{GB,Brankov,CW}). \\
Let us recall that the Curie-Weiss model for a system of $N$ spins $\sigma_{i} =\pm 1$, $i=1,\dots,N$ is defined by the Hamiltonian
\[
H = -\frac{J}{N} \sum_{i,j} \sigma_{i} \sigma_{j} + h \sum_{i} \sigma_{i}
\]
where $J$ is the spin interaction coupling constant and $h$ is the interaction constant with the external magnetic field.
The free energy is given by 
\begin{equation*}
\alpha = - \frac{1}{N} \ln Z
\end{equation*}
where the partition function
\[
Z = \sum_{\{\sigma \}} \; e^{-\beta H}
\]
is obtained by the sum over all spin configurations $\{\sigma \}$ and $\beta = T^{-1}$ where $T$ is the temperature.

Following~\cite{GB}, we introduce the notations
\[
x = - h \beta \qquad t = J \beta \qquad m = N^{-1}\sum_i^N \sigma_i \qquad \epsilon = \frac{1}{N}
\]
where $x$ and $t$ play the role of fictitious space and time variables and  $m$ is the magnetization. Hence, we can define the action
\[
S(x,t):= -\alpha = \epsilon \ln \sum_{\{ \sigma \}} \exp \left(\frac{m \; x +  m^{2} t }{\epsilon}\right).
\]
One can verify that $S(x,t)$ satisfies the potential Burgers equation
\[
S_{t} = S_{x}^{2} + \epsilon S_{xx} = \left(S_{x} + \epsilon \partial_{x} \right) S_{x}.
\]
Due to the integrability of the potential Burgers equation, it is natural to introduce the function $S(x,t_1,t_2,\dots)$ depending on an infinite set of times $t_{1},t_{2}, \dots, t_{n},\dots$ where $t_{1}:=t$ such that 
\begin{gather}
\label{rem1_potburg}
\begin{aligned}
S_{t_{n}} &= \left(S_{x}+ \epsilon \partial_{x} \right)^{n} S_{x},  \\
S(x,t) &= S(x,t,0,0,\dots),
\end{aligned}
\end{gather}
together with the initial condition 
\beq
\label{rem1_init}
S(x,0,0,\dots)  = \epsilon \ln \sum_{\{ \sigma\}}  \exp \left(  \frac{m \; x}{\epsilon} \right)  = \ln 2 + \ln \cosh x.
\eeq
It is straightforward to check that
\[
S(x,t,t_{2},\dots) = \epsilon \ln \sum_{\{ \sigma \}} \exp \left(\frac{ m \;x + m^{2} t +  m^{3} t_{2} + \dots}{\epsilon} \right)
\]
is the solution to the potential Burgers hierarchy~(\ref{rem1_potburg}) with the initial condition~(\ref{rem1_init}). We observe that the times $t_{n}$ are interpreted as the coupling constants of the $n+1-$spin interaction. Hence, the derivatives of the function $S$ with respect to $t_{n}$ provides the $n+1-$spin correlation functions 
\[
S_{t_{n}} = \langle m^{n+1} \rangle = \epsilon^{n+1} \left \langle \sum_{i_{1},i_{2},\dots,i_{n+1}} \sigma_{i_{1}} \sigma_{i_{2}} \dots \sigma_{i_{n}} \right \rangle.
\]

\end{remark}

\section{Linear viscous central invariants}

\subsection{The viscous analogue of the Camassa-Holm equation}
In the present section we show that the deformation procedure discussed above provides, in the case of the linear viscous central invariant $a(u) = u$, the following non-evolutionary conservation law
\beq\label{new}
u_t-\epsilon u_{xt}= \partial_{x} \left(u^{2} - \epsilon u u_{x} \right).
\eeq
Equation~(\ref{new}) first appeared as a scalar reduction of two-component Camassa-Holm type equations \cite{O96,Fa,LZ2}. 

This equation belongs to the class of viscous conservation laws under consideration and can be written in evolutionary form as a formal series in $\epsilon$. In fact, the application of the formal inverse operator 
\beq
\label{invop}
(1-\epsilon\d_x)^{-1}=1+\epsilon\d_x+\epsilon^2\d_x^2+...=\sum_{k=0}^{\infty}\epsilon^k\d_x^k
\eeq
to both sides of \eqref{new} gives
\begin{gather}
\begin{aligned}
\label{newexp}
u_t&=(1-\epsilon\d_x)^{-1}[2u u_x-\epsilon(u u_{xx}+u_x^2)]\\
&=(1-\epsilon\d_x)^{-1}\d_x\left[\f{u^2}{2}+(1-\epsilon\d_x)\left(\f{u^2}{2}\right)\right]\\
&=\d_x\left[\f{u^2}{2}+\sum_{k=0}^{\infty}\epsilon^k\d_x^k\left(\f{u^2}{2}\right)\right].
\end{aligned}
\end{gather}
We show that equation~(\ref{new}), or equivalently~(\ref{newexp}), can be brought to the normal form via a Miura transformation. This statement is made precise by the following
\begin{theorem}
The equation~(\ref{new}) is reduced to its normal form
\beq
\label{new2}
u_{t}=\d_x\left(\sum_k \epsilon^k u u_{(k)}\right)
\eeq
via the Miura transformation 
\beq\label{Miuranew}
u=v-\epsilon v_x.
\eeq
where $v =v(x,t)$ is a solution to the equation~(\ref{new}) or, equivalently,~(\ref{newexp}).
\end{theorem}
\proof Inverting the Miura transformation \eqref{Miuranew} we have
\[
v = (1-\epsilon \partial_{x})^{-1} u.
\]
Differentiating with respect to $t$ and using the equation~(\ref{new2}) we get
\begin{align*}
v_t &=(1-\epsilon\d_x)^{-1}u_t=(1-\epsilon\d_x)^{-1}\d_x\left(\sum_{k=0}^{\infty} \epsilon^k u u_{(k)}\right)=\d_x(1-\epsilon\d_x)^{-1}u v\\
&=\d_x(1-\epsilon\d_x)^{-1}(v^2-\epsilon v v_x)=\d_x \sum_{k=0}^{\infty} \epsilon^{k} \partial_{x}^{k} \left[v^2- \epsilon\d_x\left(\f{v^2}{2}\right)\right]\\
&=\d_x\left[v^2+\sum_{k=1}^{\infty}\epsilon^k\d_x^k\left(\f{v^2}{2}\right)\right]=\d_x\left[\f{v^2}{2}+\sum_{k=0}^{\infty}\epsilon^k\d_x^k\left(\f{v^2}{2}\right)\right]
\end{align*}
where we used the expansion formula~(\ref{invop}). The theorem is proved.
\endproof

Comparing with the results of the previous section we see that up to the fifth order in the deformation parameter the equation \eqref{new2} coincides with the equation
$$u_t=\d_x\omega^{def}_{u^2}$$
with $a(u)=u$. According with the  Conjecture \ref{maincon}, the equation~(\ref{new2}) is expected to be the only scalar integrable viscous conservation law, up to Miura transformations, possessing a linear viscous central invariant. We note that the positive hierarchy is obtained as
 the set of symmetries corresponding to the choice $f(u)=u^n,\,n=0,1,2,3,...$ (including the original equation) while the negative hierarchy
represents the set of symmetries corresponding to the choice $f(u)=u^{-n},\,n= 1, 2, 3,...$. In the next subsection we write explicitly the negative and positive hierarchies and provide an alternative proof of the integrability of equation~(\ref{new}) via linearization.
 
\subsection{The negative hierarchy}

The analysis of deformations discussed in the Section~\ref{secclass} suggests that the normal form of the equation~(\ref{new}) is not truncated. Nevertheless, there exists a family of flows in the hierarchy possessing a finite deformation, that is the negative hierarchy
\begin{equation}
\label{caln}
u_{t_{-n}}  =\partial_{x} \left( \frac{1}{u} - \epsilon \partial_{x} \frac{1}{u} \right)^{n}(\textup{1}) \qquad n=1,2,3,\dots .
\end{equation}
The first equation of this hierarchy obtained for $n=1$ 
\begin{equation}
\label{cal0}
u_{t_{-1}}  = \partial_{x} \left(\frac{1}{u} - \epsilon  \partial_{x} \frac{1}{u} \right).
\end{equation} 
has been included by F. Calogero~\cite{CAL} within a list of nonlinear PDEs that are integrable via a nonlinear change of variables.    Subsequently, it was observed in~\cite{O96} that the equation~(\ref{cal0}) can be mapped into Burgers' equation. We point out that a general family of scalar PDEs that are linearizable by a Cole-opf transformation such as the Burgers equation possessing infinitely many local symmetries has been studied in~\cite{Svinolupov}.

We also note that the hierarchy~(\ref{caln}) can be written equivalently in terms of a recursion operator as
\[
u_{t_{-n}} = R^{-1} u_{t_{-1}}
\]
where
\beq\label{invrecop}
R^{-1}=\d_x(1-\epsilon\d_x)\f{1}{u}\d_x^{-1}
\eeq
is the inverse of the operator 
\beq\label{recop}
R=\d_x u(1-\epsilon\d_x)^{-1}\d_x^{-1} 
\eeq   
such that equation~(\ref{new2}) can be written as
\beq
\label{new2rec}
u_{t} = R u_{x}.
\eeq
Operators~(\ref{invrecop}) and~(\ref{recop}) can be viewed as a reduction of the recursion operators for, respectively, negative and positive hierarchies of the system studied in \cite{Fa}. A direct proof that~(\ref{invrecop}) and~(\ref{recop}) are indeed recursion operators is based on the application on a criterion proposed by P. Olver in~\cite{O77}.
 
We now prove the integrability of equation~(\ref{new}) in two steps. First, we show that the flows \eqref{caln} can be mapped to the heat hierarchy by a sequence of hodograph and nonlinear transformations and, as a consequence, they commute. Secondly, we prove that the equation \eqref{new2} transforms into the Klein-Gordon equation and consequently commutes with the flows~(\ref{caln}). 
 
\begin{lemma}
\label{callemma}
The family of flows~(\ref{caln}) is transformed into the Burgers hierarchy via the hodograph transformation
\begin{equation}
\label{recipcal}
v(\varphi,  t_{-1},\dots,t_{-n}) = - \partial_{\varphi} x(\varphi,t_{-1},\dots,t_{-n}) \qquad \textup{where}  \qquad \varphi = \int u \; dx ,
\end{equation}
that is the function $v$ is a simultaneous solution to the Burgers hierarchy
\[
v_{t_{-n}}  = \partial_{\varphi} \left( v + \epsilon \partial_{\varphi} \right)^{n} v  \qquad n =1,2,3,\dots \;.
\]
\end{lemma}

\emph{Proof} Introducing the potential $\varphi$ such that $u = \varphi_{x}$, equations~(\ref{caln}) after the integration w.r.t. $x$ give
\begin{equation}
\label{calnpot}
\varphi_{t_{-n}} = \left(\frac{1}{\varphi_{x}} - \epsilon \partial_{x}  \frac{1}{\varphi_{x}} \right)^{n}(\textup{1}).
\end{equation}
For $n=1$ the above equation takes the form
\begin{equation}
\label{calpot}
\varphi_{t_{-1}} = \frac{1}{\varphi_{x}} + \epsilon \frac{\varphi_{xx}}{\varphi_{x}^{2}}.
\end{equation}
Introducing the hodograph transformation of the form $x = x(\varphi, t_{1},t_{2},\dots)$, one has
\begin{equation}
\label{recipder}
\varphi_{x}  = \frac{1}{x_{\varphi}} \qquad \varphi_{xx} = -\frac{x_{\varphi \varphi}}{x_{\varphi}^{3}} \qquad \varphi_{t_{n}} = - \frac{x_{t_{n}}}{x_{\varphi}}, \qquad n=1,2,\dots.
\end{equation}
Using relations above into the equation~(\ref{calpot}) one arrives to 
\begin{equation*}
x_{t_{-1}} = \left( - x_{\varphi} + \epsilon \partial_{\varphi}  \right) x_{\varphi} 
\end{equation*}
that is just the potential Burgers equation, i.e. $v =  - x_{\varphi}$ satisfies the Burgers equation
\[
v_{t_{-1}} = \partial_{\varphi} \left(v + \epsilon \partial_{\varphi} \right) v.
\]

Let us now proceed by induction assuming that the proposition is true for the $n-$th equation and consider
the $(n+1)-$th equation
\[
\varphi_{t_{-n-1}} = \left(\frac{1}{\varphi_{x}} - \epsilon \partial_{x} \frac{1}{\varphi_{x}} \right)^{n+1}(\textup{1}) =  \left(\frac{1}{\varphi_{x}} - \epsilon \partial_{x} \frac{1}{\varphi_{x}} \right) \varphi_{t_{-n}}.
\]
Using the relations~(\ref{recipder}) and observing that $\partial_{x} = x_{\varphi}^{-1} \partial_{\varphi}$, we arrive to the equation
\[
x_{t_{-n-1}} = \left(x_{\varphi} - \epsilon \partial_{\varphi} \right) x_{t_{-n}}.
\]
Since
\[
x_{t_{-n}} =  \left ( - x_{\varphi} + \epsilon \partial_{\varphi}   \right)^{n} x_{\varphi}
\]
we have
\[
x_{t_{-n-1}} =   - \left (- x_{\varphi} + \epsilon \partial_{\varphi}   \right)^{n+1} x_{\varphi}.
\]
Finally, let us differentiate the expression above by $\varphi$ and substitute $x_{\varphi} = -v$. The lemma is proved.\endproof

\begin{theorem}
The flows~(\ref{caln}) commute.
\end{theorem}

\proof In virtue of the Lemma~(\ref{callemma}) the set of flows~(\ref{caln}) is transformed into the Burgers hierarchy via the hodograph transformation~(\ref{recipcal}). On the other hand it is known that the Cole-Hopf transformation $v = \epsilon \partial_{\varphi} \log w$ brings  Burgers' hierarchy into the heat hierarchy
\[
w_{t_{-n}} = \epsilon^{n} \partial_{\varphi}^{n+1} w.
\]
This result can be straightforwardly verified for $n=1$ and then proved by induction for any $n$.
Flows of the heat hierarchy  clearly commute, i.e. $\partial_{t_{-n}} \partial_{t_{-m}} w = \partial_{t_{-m}} \partial_{t_{-n}} w = \epsilon^{n+m} \partial_{\varphi}^{n+m+2} w $. The theorem is proved.
\endproof

The following theorem uncovers the relation existing between the equation~(\ref{new}) presented above and the integrable hierarchy~(\ref{caln}).
\begin{theorem}
\label{kgth}
The flow of conservation law~(\ref{new2}) commutes with all flows of the hierarchy~(\ref{caln}).
\end{theorem} 
\proof
Let us note that  the equation \eqref{new2} can be written in the more compact form
\beq
v_t=\d_x\left[v(1-\epsilon\d_x)^{-1}v\right].
\eeq
Introducing the variable $\varphi$ defined by $\varphi_x=v$, integration with respect to $x$  provides us with the equation
\beq
\varphi_t\varphi_x -\varphi_x^3+\epsilon\left(\varphi_t\varphi_{xx}-\varphi_x \varphi_{xt} \right)=0,
\eeq
where we used the fact that the integration constant can always be eliminated by a shift of the independent variable $\varphi \to \varphi + f(t)$.
The hodograph transformation $ x = x(\varphi, t)$ brings the equation above into the form
\[
\epsilon x_{\varphi t} - x_{\varphi} x_{t} = 1,
\] 
and the change of variable $x = -\epsilon \log w$ maps the equation above into the Klein-Gordon equation in light cone variables
\[ 
\epsilon^{2} w_{\varphi t} + w = 0.
\]
Note that both the hierarchy~(\ref{caln}) and the equation~(\ref{new2}) can be linearized via the same change of variables that brings them to the heat hierarchy and Klein-Gordon equation respectively. Since the Klein-Gordon equation is compatible with all members of the heat hierarchy, the theorem is proved.
\endproof

\subsection{The positive hierarchy}
Given the equation~(\ref{new2}) in the form~(\ref{new2rec}) it is natural to consider the family of flows of the form
\beq
\label{posflows}
u_{t_{n}} = R^{n} u_{x}, \qquad n=1,2,\dots
\eeq
where $R$ is defined in~(\ref{recop}). We now prove that the family of flows~(\ref{posflows}) can be mapped into a family of linear commuting flows whose first member is the Klein-Gordon equation and for this reason we refer to it as the {\it Klein-Gordon hierarchy}.
We have the following 
\begin{theorem}
The family of flows~(\ref{posflows}) is mapped to the \textup{Klein-Gordon hierarchy} that is given by the following set of linear commuting flows
\beq
\label{kghierarchy}
w_{t_{n}} = \epsilon^{-n} \partial_{\varphi}^{-n} w, \qquad n=1,2,\dots
\eeq
via the nonlinear change of variables
\[
x(\varphi, t_1, t_{2},\dots) =  -\epsilon \log w
\]
where
\[
x = \varphi^{-1} \qquad and \qquad \varphi = \int u \; dx.
\]
Moreover, the positive hierarchy~(\ref{posflows}) and the negative hierarchy~(\ref{caln}) commute.
\end{theorem} 
\proof
Let us first note that the theorem holds true for $n=1$ in virtue of the proof of the theorem~(\ref{kgth}).
Let us now consider the equations~(\ref{posflows}) for $n\geq 2$ in the form
\[
u_{t_{n}} = R u_{t_{n-1}}, \qquad n=1,2,\dots .
\]
Introducing the function $\varphi$ such that $u = \varphi_{x}$, following the same steps as in the $n=1$ case the above equation can be equivalently written as
\[
\varphi_{x} \varphi_{t_{n}} + \epsilon \left(\varphi_{xx} \varphi_{t_{n}} - \varphi_{x} \varphi_{x t_{n}} \right) =  \varphi_{x}^{2} \varphi_{t_{n-1}}.
\]
Introducing the hodograph transformation
\[
x = x(\varphi, t_{1}, t_{2}, \dots)
\]
we arrive to the equation
\beq
\label{negburgers}
\epsilon x_{\varphi t_{n}} - x_{\varphi} x_{t_{n}} + x_{t_{n-1}} = 0.
\eeq
Setting $x=- \epsilon \log w$ one readily gets the following linear equations
\[
\epsilon w_{\varphi t_{n}} + w_{t_{n-1}} = 0, \qquad  n= 2,3,\dots
\]
that can be equivalently written in the evolutionary non-local form~(\ref{kghierarchy}). It is immediate to check the commutativity condition for the flows of both the positive and the negative hierarchy, i.e. $\partial_{t_{n}} \partial_{t_{m}} w = \partial_{t_{m}} \partial_{t_{n}} w$ for any $n,m \in \mathbb{Z} $. 
The theorem in proved.
\endproof
Summarizing, the normal form of equations associated with a linear viscous central invariant can be mapped into the Burgers equation by means of a hodograph transformation and then to the heat equation via a Cole-Hopf transformation. Interestingly, the same sequence of transformations brings on one hand the negative hierarchy into the {\it positive} Burgers hierarchy and then to the heat hierarchy. On the other hand, the positive hierarchy is mapped into the {\it negative} Burgers hierarchy whose potential form is given by the recursive nonlocal equations~(\ref{negburgers}) that is linearized  into the Klein-Gordon hierarchy (see also Figure 1)). Hence, the Klein-Gordon hierarchy can be interpreted as the negative heat hierarchy. Indeed, introducing the recursion operator
\[
{\cal R } = \epsilon \partial_{\varphi}
\]
the Burgers'  hierarchy is written as
\[
w_{t_{-n}} = {\cal R}^{n} w_{\varphi} \qquad n =0,1,2,\dots
\]
while the Klein-Gordon hierarchy is
\[
w_{t_{n}} = {\cal R}^{-n} w \qquad n =1,2, \dots .
\]
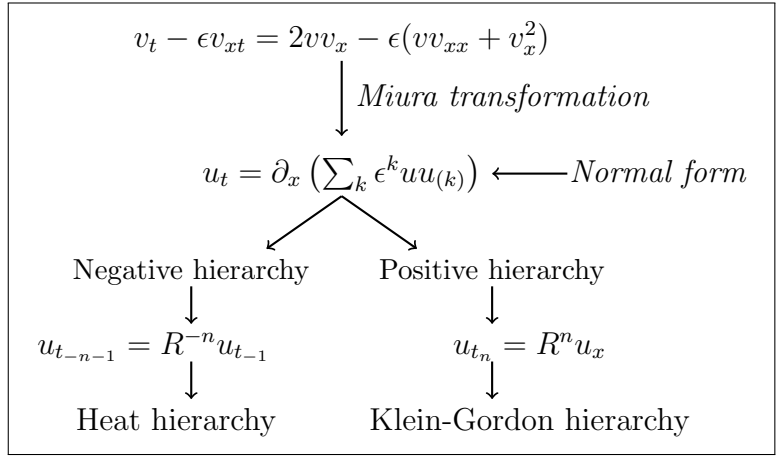
\begin{figure}
\label{scheme}
\begin{center}
\framebox{
\begin{tikzpicture}
\node at (2,1.5) {\it \; Miura transformation};
\node at (0,2.3) {$v_t-\epsilon v_{xt}=2vv_x-\epsilon(vv_{xx}+v_x^2)$};
\draw [thick, ->] (0,2) -- (0,1) ;
\node at (0,.5) {$u_{t}=\d_x\left(\sum_k \epsilon^k u u_{(k)}\right)$};
\draw [thick,<-] (2,.5) -- (3,.5) ;
\node at (4.2,.5) {\it Normal form};
\draw [thick,->] (0,.2) -- (-1,-.5) ;
\node at (-2,-.8) {{\small Negative hierarchy}};
\node at (2,-.8) {{\small  Positive hierarchy}};
\draw [thick,->] (-2,-1) -- (-2,-1.5) ;
\draw [thick,->] (0,.2) -- (1,-0.5) ;
\draw [thick,->] (2,-1) -- (2,-1.5) ;
\node at (-2.5,-1.8) {$u_{t_{-n-1}}  = R^{-n} u_{t_{-1}} $};
\node at (2.5,-1.8) {$u_{t_n}=R^{n}u_x $};
\draw [thick,->] (-2,-2) -- (-2,-2.5) ;
\draw [thick,->] (2,-2) -- (2,-2.5) ;
\node at (-2.2,-2.8) {Heat hierarchy };
\node at (2.5,-2.8) {Klein-Gordon hierarchy};
\end{tikzpicture}
}
\caption{Positive and negative integrable hierarchies associated with viscous conservation laws possessing a linear viscous central invariant.}
\end{center}
\end{figure}

\section{Local well-posedness of the Cauchy problem}

\subsection{Kato's method}

The present Section is devoted to the study of the local (in time) well-posedness of the periodic Cauchy problem for \eqref{new} by using Kato's method \cite{Kato}  that is here briefly outlined, following the statement which is theorem 1 in \cite{Constantin}.

Let us consider the Cauchy problem for the quasi-linear equation of evolution: 
\beq\label{general}
\frac{d}{dt}v+A(v)v=f(v), \quad t>0\quad 
v(0)=\phi.
\eeq
Let $X$ and $Y$ be reflexive Banach spaces with $Y$ continuously and densely embedded in $X$, let $S: Y \to X$ be an isomorphism and suppose that the norm of $Y$ is chosen in such a way that $S$ is an isometry. Assume now that the terms appearing in \eqref{general} satisfy the following conditions:
\begin{enumerate}
\item $A$ is an operator-valued function defined on $Y$, that is for each $y\in Y$, $A(y): D(A)\subset X \rightarrow  X$ is a linear operator on $X$ (in the interesting cases $A(y)$ is an unbounded operator). Assume that $A(y)$ is quasi-m-accretive, uniformly for $y\in Y$ with $\|y\|_Y\leq M$. In other words, for every constant $M>0$, there is a real number $\omega$ such that for every $y\in Y$, with $\|y\|_Y\leq M$, $-A(y)$ generates a $C^0$-semigroup $\{e^{-tA(y)}\}_{t\geq 0}$ with 
$$\|e^{-tA(y)}\|_{\mathcal{L}(X)}\leq e^{\omega t}, \quad t\geq 0,$$
where $\|\cdot\|_{\mathcal{L}(X)}$ is the operator norm.
\item For each $y\in Y$, $A(y)$ is a bounded linear operator from $Y$ to $X$ and moreover
\begin{equation}\label{point2} \|(A(y)-A(z))w\|_X\leq \mu_A \|y-z\|_X\|w\|_Y, \quad y,z, w\in Y, \end{equation}
for some constant $\mu_A$ depending only on $\max\{\|y\|_Y, \|z\|_Y\}$. 
\item for any $M>0$, the inequality 
\begin{equation}\label{point3}\|(SA(y)-A(y)S)S^{-1}w\|_X\leq \mu_1(M)\|w\|_X, \quad y\in Y, \, \|y\|_Y\leq M\end{equation}
holds for all $w\in Y$, where $\mu_1(M)>0$ is a constant. 
\item For each $M>0$, $f$ is a bounded function from $\{y \in Y: \, \|y\|_Y\leq M\}$ to $Y$. Also we have 
\begin{equation}\label{point41}\|f(y)-f(z)\|_X\leq \mu_2 \|y-z\|_X, \quad y, z\in Y,\end{equation}
and 
\begin{equation}\label{point42}\|f(y)-f(z)\|_Y\leq \mu_3\|y-z\|_Y, \quad y, z\in Y,\end{equation}
for some constants $\mu_2$ and $\mu_3$, where $\mu_2$ depends only on $\max\{\|y\|_{Z}, \|z\|_{Z}\}$ and $\mu_3$ depends only on $\max\{\|y\|_{Y}, \|z\|_{Y}\}, $ where $Z$ is a space such that $Y\subset Z\subset X$, all with continuous inclusions. (In our application it will be $X:=L^2, Y:=H^1, Z:=L^{\infty}$ all for functions defined on the circle.)\footnote{Note that these assumptions are exactly like those stated in \cite{Constantin}, except for the Lipschitz inequality \eqref{point41}. Indeed in \cite{Constantin} it is required that $\mu_2$ depends only on $\max\{\|y\|_{X}, \|z\|_{X}\}$, while here we require dependence on a bigger quantity. As it can be seen in \cite{Kato} page 40, second inequality from above, this is not affecting the statement of Theorem \ref{kato} in our case.} 
\end{enumerate}
Then one has the following Theorem (see \cite{Kato}):
\begin{theorem}\label{kato}
Assume conditions (1), (2), (3) and (4) above hold. Then for any $\phi\in Y$, there is a $T>0$, depending only on $\|\phi\|_Y$ and a unique solution $v$ to \eqref{general} such that $v\in C^0([0, T], Y)\cap C^1([0, T], X)$. Moreover, $v(t)$ depends continuously on the initial data $\phi=v(0)$ in the $Y$-norm. 
\end{theorem}

\subsection{The periodic Cauchy problem for the equation~(\ref{new})}

Let us consider the equation~(\ref{new}) with fixed $\epsilon$ and, for convenience, we set $\epsilon = 1$ and replace $t \to -t$. With these notations the initial value problem for the equation~(\ref{new}) will read as follows
\beq
\label{newm}
m_t=-v m_x-v_x m \qquad v(x,0) = w(x)
\eeq
where $m := v - v_{x}$ and $w: \mathbb{S}^{1} \to \mathbb{R}$. The above time inversion will result into proving the local well-posedness of the Cauchy problem for equation~(\ref{new}) backward in time. We will finally observe that the result holds also true for the forward evolution. 

Using the fact that $v=S^{-1}m$  where $S = 1-\partial_{x}$  the problem~(\ref{newm}) can be equivalently formulated as follows
\beq\label{periodicCauchy}
m_t=-(S^{-1}m) m_x-m(S^{-1}m)_x, \quad
m(x,0)=\phi(x)
\eeq
where  $\phi: \mathbb{S}^{1} \to \mathbb{R}$.
Verifying the conditions for the validity of Kato's theorem will guarantee  well-posedness for $\phi\in H^1(\mathbb{S}^1)$.

From now on all the functional spaces introduced are referred to functions on the circle and we for sake of simplicity will be denoted just as $L^2$, $H^1$. 

We have the following 
\begin{theorem}\label{localexistence}
Let $\phi\in H^1$, then there is a $T>0$, depending only on $\|\phi\|_{H^1}$ such that \eqref{periodicCauchy} has a unique solution $m\in C^0([0, T], H^1)\cap C^1([0,T], L^2)$ and moreover $m$ as a function of $t$ depends continuously on $\phi$ in the $H^1$-norm. 
\end{theorem}
\proof
We have just to check assumptions (1) to (4) of Theorem \ref{kato}. We choose as $X=L^2$ and as $Y=H^1$. On $L^2$ we use the usual $L^2$-norm.  Recall that $H^1$ is continuously and densely embedded in $L^2$, since for any $y\in H^1$, $\|y\|_{L^2}\leq \|y\|_{H^1}$ (by definition $\|y\|^2_{H^1}:=\|y\|^2_{L^2}+\|y_x\|^2_{L^2}$). Also we choose as an isometric isomorphism $S=1-\partial_x$, so that 
$$\|Sy\|^2_X=\|y-y_x\|^2_{L^2}=<y-y_x, y-y_x>_{L^2}=$$
$$=\|y\|^2_{L^2}+ \|y_x\|^2_{L^2}-<y_x, y>_{L^2}-<y, y_x>_{L^2},$$
where the terms $<y_x, y>_{L^2}+<y, y_x>_{L^2}$ give a zero contribution due to integration by parts which holds for functions in $H^1$ (see for instance Corollary 8.1 \cite{Brezis}). Therefore, with the choice of $S$ as an isometric isomorphism, the norm on $H^1$ is equal to the standard norm on $H^1$. Comparing \eqref{periodicCauchy} with \eqref{general}, we choose 
$$A(y)=(S^{-1}y)\partial_x, \quad f(y)=-y(S^{-1}y)_x, \quad y\in Y=H^1.$$
We start verifying assumption (1), fixing $M>0$ and $y\in Y$ with $\|y\|_Y\leq M$. 
First notice that $(S^{-1}y)\in Y$ but since  
\beq\label{smoothness}
(S^{-1}y)_x=S^{-1}y-y
\eeq 
it turns out that $S^{-1}y$ has actually a $C^1$ representative (this is because $H^1$ is embedded in $C^0$). In order to check condition (1) then, we need just to show that 
$\sup_{x\in [0,1]}|(S^{-1}y)_x|$ is bounded by a constant depending only on $M$ (see \cite{Kato}, p. 38). On the other hand, by \eqref{smoothness}
$$\sup_{x\in [0,1]}|(S^{-1}y)_x|\leq \|S^{-1}y\|_{L^{\infty}}+\|y\|_{L^{\infty}}\leq c\|S^{-1}y\|_{H^1}+c\|y\|_{H^1},$$
where $c$ is the constant entering in the Sobolev inequality corresponding to the inclusion $H^1\hookrightarrow L^{\infty}$. Since $S$ is an isometry we have $\|S^{-1}y\|_{H^1}=\|y\|_{L^2}$ and $\|y\|_{L^2}\leq \|y\|_{H^1}$ from which we deduce
\beq\label{bound1}\sup_{x\in [0,1]}|(S^{-1}y)_x|\leq 2cM.\eeq
Therefore, $A$ is quasi-m-accretive on $L^2$ with $\omega$ bounded by $cM$. 

To check assumption (2) we first show that $A(y)$ is a bounded operator from $H^1$ to $L^2$ for every $y\in H^1$. 
Indeed we have, for $w\in H^1$: $$\|A(y)w\|_{L^2}=\|(S^{-1}y)w_x\|_{L^2}\leq \|S^{-1}y\|_{L^{\infty}}\|w_x\|_{L^2}\leq c \|S^{-1}y\|_{H^1}\|w\|_{H^1}=
$$
$$= c\|y\|_{L^2}\|w\|_{H^1}\leq c\|y\|_{H^1}\|w\|_{H^1}.$$
To conclude the verification of (2), namely showing that \eqref{point2} is fulfilled, we just observe that 
$$\|(A(y)-A(z))w\|_{L^2}=\|S^{-1}(y-z)w_x\|_{L^2}\leq\|S^{-1}(y-z)\|_{L^{\infty}}\|w_x\|_{L^2}\leq$$
$$\leq c\|S^{-1}(y-z)\|_{H^1}\|w\|_{H^1}=c\|y-z\|_{L^2} \|w\|_{H^1}.$$
Let us now verify the estimate \eqref{point3}. We notice that 
$$SA(y)v-A(y)Sv=-(S^{-1}y)_x v_x,$$
then, for all $y, w\in H^1$, with $\|y\|_{H^1}\leq M$ we have  $$\|(SA(y)-A(y)S)S^{-1}w\|_{L^2}=\|-(S^{-1}y)_x(S^{-1}w)_x\|_{L^2},$$
which in virtue of \eqref{smoothness} becomes $$\|(S^{-1}y-y)(S^{-1}w-w)\|_{L^2}.$$
Therefore, $$\|(SA(y)-A(y)S)S^{-1}w\|_{L^2}=\|(S^{-1}y)(S^{-1}w)+yw-(S^{-1}w)y-(S^{-1}y)w\|_{L^2}\leq $$
$$\leq \|(S^{-1}w)(S^{-1}y-y)\|_{L^2}+\|w(S^{-1}y-y)\|_{L^2}\leq \left(\|S^{-1}w\|_{L^2}+\|w\|_{L^2}\right)\left(\|S^{-1}y\|_{L^{\infty}}+\|y\|_{L^{\infty}}\right).$$
Since $\|S^{-1}y\|_{L^{\infty}}\leq c\|S^{-1}y\|_{H^1}=c\|y\|_{L^2}\leq c\|y\|_{H^1},$
$\|y\|_{L^{\infty}}\leq c\|y\|_{H^1}$ and $\|S^{-1}w\|_{L^2}\leq \|S^{-1}y\|_{H^1}=\|y\|_{L^2}$, we get
$$\|(SA(y)-A(y)S)S^{-1}w\|_{L^2}\leq 4c \|y\|_{H^1}\|w\|_{L^2}\leq 4cM\|w\|_{L^2}.$$

In order to check  the assumption (4), we first show $f(y):=-y(S^{-1}y)_x$ is bounded from $L:=\{y\in H^1: \|y\|_{H^1}\leq M\}$ to $H^1$. 
We have 
$$\|y(S^{-1}y)_x\|^2_{H^1}=\|y(S^{-1}y)_x\|^2_{L^2}+\|\partial_x\left(y(S^{-1}y)_x \right)\|^2_{L^2},$$
and 
$$\|y(S^{-1}y)_x\|^2_{L^2}\leq \|(S^{-1}y)_x\|^2_{L^{\infty}}\|y\|_{L^2}^2\leq4c^2M^4,$$
in virtue of \eqref{bound1} and $\|y\|_{L^2}\leq \|y\|_{H^1}\leq M$. 
To control the other term, we observe that 
$$\|\partial_x\left(y(S^{-1}y)_x \right)\|_{L^2}\leq\|y_x(S^{-1}y)_x\|_{L^2}+\|y(S^{-1}y)_{xx}\|_{L^2}\leq $$
$$\leq \|(S^{-1}y)_x\|_{L^{\infty}}\|y_x\|_{L^2}+\|y(S^{-1}y-y-y_x)\|_{L^2}\leq 2cM^2+\|y\|_{L^{\infty}}(\|S^{-1}y\|_{L^2}+\|y\|_{L^2}+\|y_x\|_{L^2})\leq$$
$$\leq 2cM^2+cM(3M)=5cM^2, $$
where we have used once again \eqref{bound1} and $\|y_x\|_{L^2}\leq \|y\|_{H^1}\leq M$. 
Combining these estimates we obtain 
$$\|y(S^{-1}y)_x\|^2_{H^1}\leq 4c^2 M^4+25 c^2 M^4=29 c^2 M^4.$$

\noindent Let us prove the first Lipschitz inequality \eqref{point41} is fulfilled. We have
$$\|f(y)-f(z)\|_{L^2}\leq \|yS^{-1}y-zS^{-1}z\|_{L^2}+\|y^2-z^2\|_{L^2}.$$
For the first term $\|yS^{-1}y-zS^{-1}z\|_{L^2}$, adding and subtracting $zS^{-1}y$ we obtain 
\begin{align}
\label{ineq10}
&\|yS^{-1}y-zS^{-1}z\|_{L^2}\leq \|S^{-1}y\|_{L^{\infty}}\|y-z\|_{L^2}+\|S^{-1}(y-z)\|_{L^{\infty}}\|z\|_{L^2} \leq\\
\label{ineq11}
&\leq \|S^{-1}y\|_{H^1}\|y-z\|_{L^2}+\|S^{-1}(y-z)\|_{H^1}\|z\|_{L^2}=(\|y\|_{L^2}+\|z\|_{L^2})\|y-z\|_{L^2} \\ 
&\leq \mu\|y-z\|_{L^2},
\end{align}
where $\mu=\max\{\|y\|_{L^2}, \|z\|_{L^2}\}$. 
Notice now that we can not bound the term $\|y^2-z^2\|_{L^2}$ with a Lipschitz constant depending only on the norms $\|y\|_{L^2}$ and $\|z\|_{L^2}$, but we will have to use the $L^{\infty}$-norms that are bigger. This is still sufficient, as one can see in the second inequality at page 40 of \cite{Kato}.
Then it is immediate to get \beq\label{ineq100}\|y^2-z^2\|_{L^2}\leq(\|y\|_{L^{\infty}}+\|z\|_{L^{\infty}}) \|y-z\|_{L^2}.\eeq Combining this with the inequality controlling $ \|yS^{-1}y-zS^{-1}z\|_{L^2}$, \eqref{point41} is checked. 

Finally we prove the last Lipschitz inequality \eqref{point42}. 
We have for every $y, z\in H^1$ 
\beq\label{ineq50}\|y(S^{-1}y)_x-z(S^{-1}z)_x\|_{H^1}\leq \|y(S^{-1}y)-z(S^{-1}z)\|_{H^1}+\|y^2-z^2\|_{H^1},\eeq
using \eqref{smoothness}. 
We first control the first term on the right: 
$$\|y(S^{-1}y)-z(S^{-1}z)\|_{H^1}\leq \|y(S^{-1}y)-z(S^{-1}z)\|_{L^2}+\|\partial_x\left( y(S^{-1}y)-z(S^{-1}z)\right)\|_{L^2}\leq $$
$$\leq \|y(S^{-1}y)-z(S^{-1}z)\|_{L^2}+\|y_x(S^{-1}y)-z_x(S^{-1}z)\|_{L^2}+\|y(S^{-1}y)_x-z(S^{-1}z)_x
\|_{L^2}\leq$$
$$\leq (\|y\|_{L^2}+\|z\|_{L^2})\|y-z\|_{L^2}+\|y_x(S^{-1}y)-z_x(S^{-1}z)\|_{L^2}+$$
$$+(\|y\|_{L^2}+\|z\|_{L^2})\|y-z\|_{L^2}+(\|y\|_{L^{\infty}}+\|z\|_{L^{\infty}})\|y-z\|_{L^2},$$
using the inequalities \eqref{ineq10}, \eqref{ineq11} and \eqref{ineq100}. 
Therefore we obtain for the first term on the right hand side of \eqref{ineq50}
$$\|y(S^{-1}y)-z(S^{-1}z)\|_{H^1}\leq 3(\|y\|_{L^{\infty}}+\|z\|_{L^{\infty}})\|y-z\|_{H^1}+\|y_x(S^{-1}y)-z_x(S^{-1}z)\|_{L^2}.$$
On the other hand, adding and subtracting $y_x(S^{-1}z)$ inside $\|y_x(S^{-1}y)-z_x(S^{-1}z)\|_{L^2}$ we obtain 
$$\|y_x(S^{-1}y)-z_x(S^{-1}z)\|_{L^2}\leq \|y_x(S^{-1}(y-z))\|_{L^2}+\|(S^{-1}z)(y_x-z_x)\|_{L^2}\leq$$
$$\leq \|S^{-1}(y-z)\|_{L^{\infty}}\|y_x\|_{L^2}+\|S^{-1}z\|_{L^{\infty}}\|y_x-z_x\|_{L^2}\leq $$
$$\leq c\|S^{-1}(y-z)\|_{H^1}\|y\|_{H^1}+c\|S^{-1}z\|_{H^1}\|y-z\|_{H^1}\leq c\left( \|y\|_{H^1}+\|z\|_{H^1}\right)\|y-z\|_{H^1}.$$
This proves that the first term on the right hand side of \eqref{ineq50} is bounded by $\mu\|y-z\|_{H^1}$ where $\mu$ depends only on $\max\{\|y\|_{H^1}, \|z\|_{H^1}\}$. 
Finally to control the second term on the right hand side of \eqref{ineq50} observe that 
$$\|y^2-z^2\|_{H^1}\leq \|(y+z)(y-z)\|_{L^2}+\|\partial_x\left((y-z)(y+z)\right)\|_{L^2}\leq$$
$$\leq \|(y+z)(y-z)\|_{L^2}+\|(y-z)_x(y+z)\|_{L^2}+\|(y-z)(y+z)_x\|_{L^2}\leq $$
$$\leq \|y+z\|_{L^{\infty}}\|y-z\|_{L^2}+\|(y-z)_x\|_{L^2}\|y+z\|_{L^{\infty}}+\|y-z\|_{L^{\infty}}\|(y+z)_x\|_{L^2}\leq $$
$$\leq c(\|y\|_{H^1}+\|z\|_{H^1})\|y-z\|_{H^1}+c\|y-z\|_{H^1}(\|y\|_{H^1}+\|z\|_{H^1})+c\|y-z\|_{H^1}(\|y\|_{H^1}+\|z\|_{H^1})\leq $$
$$\leq 3c (\|y\|_{H^1}+\|z\|_{H^1})\|y-z\|_{H^1}.$$
Therefore we can conclude that 
$$\|y(S^{-1}y)_x-z(S^{-1}z)_x\|_{H^1}\leq\mu_3 \|y-z\|_{H^1},$$
where $\mu_3$ depends only on $\max\{\|y\|_{H^1}, \|z\|_{H^1}\}$ so assumption (4) is satisfied and the Theorem is proved. 
\endproof
\begin{remark}
In general the time $T$ in the Theorem \ref{localexistence} depends on the norm $\|\phi\|_{H^1}$, hence for each choice of $M>0$ such that $\|\phi\|_{H^1}\leq M$ we will have a corresponding $T$ depending on $M$ (see \cite{Kato}). 
Moreover,  since the operator $A(y)$ used in the above proof generates a $C^0$-semigroup, and not just $-A(y)$ (see for instance \cite{RR}, Theorem 12.26) we have local well-posedness also forward in time for the original equation \eqref{new}. 
\end{remark}

\section{Numerical solutions}
\begin{figure}  
\centering
\includegraphics[width=15cm]{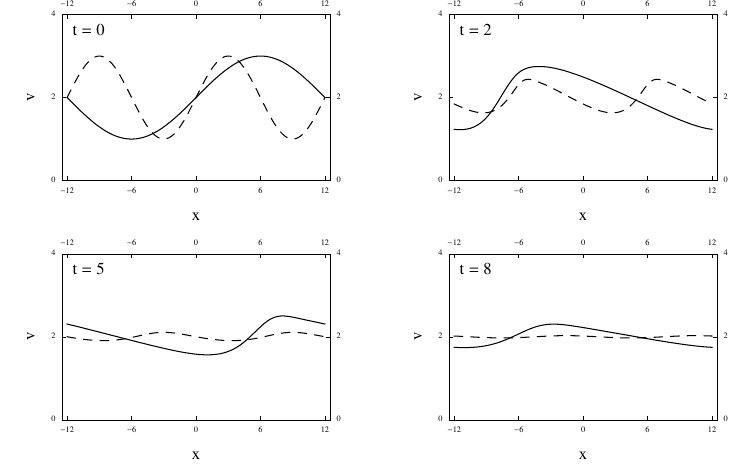}  
\caption{Time evolution for the initial datum $v(x,0)=\sin\left(\frac{\pi x}{12}\right)+2$ (solid line) and $v(x,0) = \sin \left (\frac{\pi x}{6} \right) + 2$ (dashed line) with periodic boundary conditions.}  
\label{viscon1}
\end{figure}

The present section is devoted to the numerical study of the initial value problem for the equation~\eqref{new} in the case of periodic boundary conditions. Let us observe that introducing the auxiliary variable    $P(x,t)$ such that
\beq
\label{Pdef}
(1-\p_x)P=\frac{1}{2}v^2,
\eeq
equation~(\ref{new}) can be equivalently written as follows
\beq
\label{newP}
(1-\p_x)\left(v_t-v v_x -P_x\right)=0.
\eeq
Since the only smooth function $f$ fulfilling periodic boundary conditions such that $(1 -\p_x)f=0$ is $f(x,t) \equiv 0$, the equation~\eqref{newP} is equivalent (on smooth solutions) to the system 
\beq
\label{inter2}
\begin{split}v_t=& v v_x+P_x, \quad \\ P_x=& P-\frac{1}{2} v^2.
\end{split}
\eeq
Numerical solutions are obtained by using a pseudo spectral method implemented in the {\tt NDSolve Package} of {\tt Mathematica 9} \cite{W}. Although  local existence of solutions is guaranteed by the Theorem~{\ref{localexistence}}, we observe that positive definite initial data seems to support global existence (in time) of solutions. On the contrary, finite-time break up occurs in other cases. 
In Figure~{\ref{viscon1}} we compare the evolution at different time steps of the positive definite sinusoidal initial data 
\begin{align}
\label{initv1}
v_{1}(x) := & v(x,0) =\sin \left(\frac{\pi x}{12} \right) + 2 \\
\label{initv2}
v_{2}(x) := & v(x,0) =\sin \left(\frac{\pi x}{6} \right) + 2.
\end{align}
From the definition of $P(x,t)$ we have respectively
\begin{align*}
P_{1}(x) := &P(x,0)=\frac{9}{4}-\,{\frac {9 \cos \left( \frac{1}{6}\pi x \right) }
{36+{\pi}^{2}}}+{\frac {3 \pi\,\sin \left( \pi x
 \right) }{2 (36+{\pi}^{2})}}+\frac{24 \pi \cos \left( \frac{1}{12}\pi x
 \right) } { 144+\,{\pi}^{2} }+\frac{288\sin
 \left( \frac{1}{12}\pi x \right) } { 144+ {\pi}^{2}} \\
P_{2}(x) :=& P(x,0)= \frac{12 \pi \cos \left (\frac{\pi x}{6} \right )}{\pi^{2}+36} - \frac{9 \cos \left (\frac{\pi x}{3} \right)}{4 (\pi^{2} + 9)} + \frac{72 \sin \left (\frac{\pi x}{6} \right)}{\pi^{2} + 36} + \frac{3 \pi \sin \left (\frac{\pi x}{3} \right)}{9 (\pi^{2} + 9)} + \frac{9}{4}.
\end{align*}
In both cases, after the initial steepening both solutions tend to vanish in time. It should be noted that, as expected, the steepening is more pronounced for  the datum of larger period~(\ref{initv1}), although the effect of dissipation prevents it to reach the gradient catastrophe. In the case of the Camassa-Holm equation it was proved that if  the quantity $u(x,0)-u_{xx}(x,0)$ does not change sign, the periodic Cauchy problem is globally well-posed in time in $H^2$ \cite{Constantin}. It would be interesting to prove a similar results for the equation~(\ref{new}).
 
\begin{figure}
\centering  
\includegraphics[width=15cm]{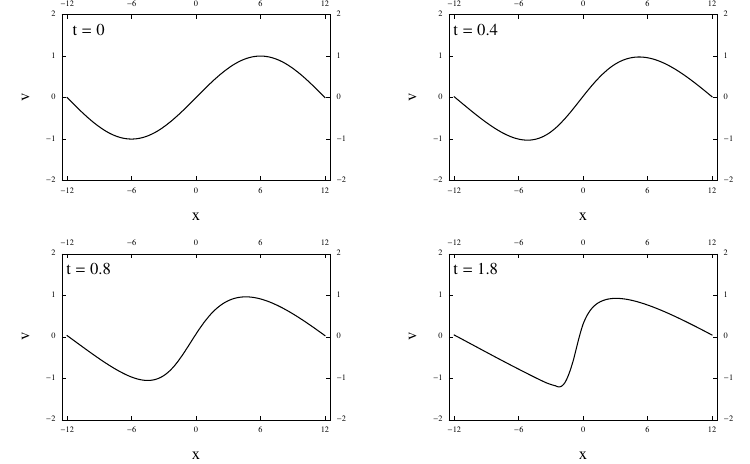}  
\label{fig:fig6}
\caption{Steepening of the solution for the initial datum $v(x,0)=\sin \left (\frac{\pi x}{12} \right)$ .}  
\label{viscon2}
\end{figure}

A completely different scenario takes place when considering initial data that change sign, as for instance an odd initial datum as
\[
v_{3}(x) := v(x,0)=\sin \left( \frac{\pi x}{12} \right).
\] 
The corresponding initial function $P$ is
\begin{align*}
P_{3}(x) := P(x,0) = \frac{3\pi \sin \left(\frac{\pi x}{6} \right)}{2 (\pi^{2} + 36)} - \frac{9 \cos \left(\frac{\pi x}{6} \right)}{\pi^{2} + 36} + \frac{1}{4}.
\end{align*}
As Figure~\ref{viscon2} shows no damping occurs in this case and the solution seems to develop a gradient catastrophe in finite time at the inflection point $x=0$. 
A similar phenomenon has been observed in the case of Camassa-Holm equation for which there exists a class of initial conditions ($u(x,0)\in H^4$ with $u(x,0)$ odd and $u(x,0)\neq 0$) such that the solution blows up in finite time \cite{Constantin}.
 In the case of Cauchy problem on the real line, the existence of solutions of Camassa-Holm equation with finite-time blow-up was previously argued in \cite{CH}.


\begin{remark}
Let us observe that if $G$ is the Green function for the operator $(1-\p_x)$ on the circle, then from the definition of the funtion $P$~\textup{(\ref{Pdef})} we have 
$P=G\star \frac{1}{2}v^2$, where $\star$ denotes the convolution product. In full analogy with the Camassa-Holm equation,  using this representation for $P$ the equation \eqref{new} can be written as a conservation law with nonlocal flux
$$v_t=\p_x\left(\frac{1}{2}v^2+G\star \frac{1}{2}v^2 \right).$$

\end{remark}

\section{Quasitriviality, transport equations and deformed hodograph formula}
A Miura transformation is defined as a change of variables of the form
\beq
\label{qM}
u(v,v_x,...)=\sum_k\epsilon^k F_k(v,v_x,v_{xx},...),\qquad {\rm deg}F_k=k
\eeq
where coefficients $F_k$ are differential polynomials in the derivatives $v_x,v_{xx},..$. A natural generalization of~(\ref{qM}), referred to as  \emph{quasi-Miura} transformation, is obtained just relaxing the assumption on the polynomial form of $F_{k}$.

According to the results of \cite{LZ}, any evolutionary PDE of the form
$$v_t=vv_x+\sum_k \epsilon^k P_k(v,v_x,v_{xx},...),\qquad {\rm deg}P_k=k$$
can be reduced via a \emph{quasi-Miura} transformation to the Hopf equation~(\ref{hopf}).
 Moreover any evolutionary
 symmetry of the equation
$$v_{\tau}=c(v)v_x+\sum_k \epsilon^k Q_k(v,v_x,v_{xx},...)$$
 is reduced by the same transformation to its dispersionless limit
$$u_{\tau}=c(u)u_x.$$

In particular, given an integrable hierarchy of evolutionary PDEs, there exists an invertible quasi-Miura transformation that brings any equation of the hierarchy to the corresponding equation of the Hopf hierarchy.

The method for constructing recursively the terms of  a quasi-Miura transformation at any order in $\epsilon$ has been illustrated by S. Liu and Y. Zhang in~\cite{LZ} and it is based on the construction of  infinitesimal generators $X_1,X_2,...$ for the reducing transformation 
\beq
v(u)=\exp{\tilde{X}}(u)=u+\tilde{X}(u)+\f{1}{2}\tilde{X}\left(\tilde{X}(u)\right)+�
\eeq
where 
$$\tilde{X}=X\f{\d}{\d u}+X_x\f{\d}{\d u_x}+...$$
and
$$X=\epsilon X_1+\epsilon^2 X_2+.... \qquad .$$
A similar result involving a more general class of transformations that allow also a dependence on $x$ and $t$ was obtained in \cite{BGI}. 
The method illustrated in~\cite{BGI} is based on the construction of the transformation that reduces the perturbed equation to the unperturbed one the via the solution of the transport equations.

Hence, inspired by~\cite{BGI}, we provide a straightforward proof of the quasi-triviality for the class of equations of the form
\beq
\label{claw}
v_{t} = \partial_{x} \left(v^{2} + \epsilon a(v) v_{x} + \epsilon^{2} b_{1}(v) v_{xx} + \epsilon^{3} c_{1}(v) v_{xxx} + \dots \right).
\eeq
Moreover, exploiting the additional freedom provided
 by the solutions of transport  equations we show how to construct a generalized quasi-Miura transformation that preserves the initial datum.
\begin{theorem}
Scalar conservation laws~(\ref{claw}) are quasitrivial, i.e. they can be reduced to the Hopf equation by a quasi-Miura tranformation.
\end{theorem}
\proof
Looking for asymptotic solutions in power series of $\epsilon$ of the form
\beq
\label{formalsol}
v = u + \epsilon v^{1} + \epsilon^{2} v^{2} + \dots
\eeq
where $u = v^{0}$, the equation~(\ref{claw}) splits into a quasilinear equation for $u_{0}$ plus a set of transport equations
\begin{gather}
\label{transport}
\begin{aligned}
&L u = 0 \\
&L^{*} v^{1} = -\partial_{x} \left(a(u) u_{x} \right) \\
&L^{*} v^{2} =-\partial_{x} \left( (v^{1} + \partial_{x} a(u)) v^{1} + b_{1}(u) u_{xx} \right)\\
&\dots
\end{aligned}
\end{gather}
where
\[
L = \partial_{t} - 2 u  \partial_{x} \qquad  L^{*} =- \partial_{t} + \partial_{x} 2 u 
\]
are, respectively, the first order differential operator of the Hopf flow and its formal adjoint. Hence, the problem of constructing the reducing quasi-Miura transformation is reduced to the classical problem of constructing the expansion of the form~(\ref{formalsol}) via the transport equations~(\ref{transport}). Such solutions  are referred to as formal solutions to the equation~(\ref{claw}). Let us observe that once a solution $u$ to the first equation in~(\ref{transport}) is known, higher order corrections $v^{1}$, $v^{2}$, $\dots$ are obtained by solving a sequence of linear PDEs with coefficients depending on $u$. Hence,  the asymptotic formal solution of~(\ref{formalsol}) can be interpreted as a transformation that brings any solution to the Hopf equation to a solution of the deformed equation~(\ref{claw})\cite{DZ,BGI}. It turns out that for the class of equations under consideration such a transformation is of quasi-Miura type.
Given a solution to the Hopf equation $L u = 0$ that is implicitly given in terms of hodograph equation:
\begin{equation}
\label{Lusol}
x + 2 u t + f(u) = 0,
\end{equation}
the $n$-th transport equation can be written in the form
\begin{equation}
\label{Lvn}
L^{*} v^{n}  =F_{n}\left(u,u_{x} \right),
\end{equation}
where we have observed that  using the differential consequences of the equation~(\ref{Lusol})
\begin{equation}
\label{triang}
u_{xx} = f''(u) u_{x}^{3}, \qquad u_{xxx} = f'''(u) u_{x}^{4} + 3 (f'')^{2} u_{x}^{5} \qquad \textup{etc.}
\end{equation}
the r.h.s can be written as a suitable function of $u$ and $u_{x}$ only. The general integral of the equation~(\ref{Lvn}) is readly obtained as a function of the variables $u$ and $u_{x}$ of the form
\begin{equation}
\label{Lvnint}
v^{n} = p_{n}(u,u_{x}) u_{x}  + h_{n}(u, u_{x})
\end{equation}
where
\[
p_{n}(u,u_{x}) = \int^{u_{x}} \frac{1}{2 \varphi^{3}} F_{n}(u,\varphi)\; d\varphi
\]
and $h_{n}(u,u_{x})$ is the general solution to the homogenous linear equation $L^{*}h_{n} = 0$. Using the method of characteristics one can show that
\[
h_{n}(u,u_{x})= g_{n}(u) u_{x} 
\]
where $g_{n}(u)$ is an arbitrary function of its argument. 
\endproof
For the Burgers equation, that is obtained from~(\ref{claw}) by choosing a constant central invariant $a(u) = 1$ the set of transport equation takes the a simple recursive form
\begin{gather}
\label{btransport}
\begin{aligned}
&L^{*}v^{1} =- \partial_{x} (u_{x}) \\
&L^{*}v^{N} =- \partial_{x} \left( \sum_{i = 1}^{N-1} v^{i} v^{N-1}  + v^{N-1}_{x}   \right), \qquad N = 2,3,\dots
\end{aligned}
\end{gather}
with the notation $v^{0} = u$. 
\begin{theorem}
The solution to the $n-$th transport equation for the Burgers equation is given by
\begin{equation}
\label{tburgers}
v^{n}  = \sum_{j=1}^{2n-1} \alpha_{n,n+j} \; u_{x}^{n+j} + g_{n}(u) u_{x}
\end{equation}
where coefficients $\alpha$ are determined by recursion as follows
\begin{align*}
\alpha_{n,n+1} =& \frac{1}{2n} \alpha''_{n-1,n}  \\
\alpha_{n,n+2} =& \frac{1}{2 (n+1)}  \left [\alpha''_{n-1,n+1}+ \Lambda'_{n-1,2}  + \Omega_{n-1,1}\right ] \\
\alpha_{n,n+j} =&\frac{1}{2(n+j-1)} \left[ \alpha''_{n-1,n+j-1} + \Lambda'_{n-1,j} + (n+j-1) f''(u) \Lambda_{n-1,j-1} + \Omega_{n-1,j-1} \right . \\ 
&\left .+ (n+j-3) (n+j-1)( f''(u))^{2} \alpha_{n-1,n+j-3} \right ], \qquad  j =3,\dots,2n-3 \\
\alpha_{n,3n-2} =& \frac{1}{2(3n-3)} \left[\Lambda'_{n-1,2n-2} + (3n-3) f''(u) \Lambda_{n-1,2n-3} + \Omega_{n-1,2n-3} \right .\\
&\left .+ (3n-5) (3n-3) (f''(u))^{2} \alpha_{n-1,3n-5}  \right]  \\
\alpha_{n,3n-1} =& \frac{1}{2 (3n-2)} \left [(3n-2) f''(u) \Lambda_{n-1,2n-2} + 3 n (f''(u))^{2} \alpha_{n-1,3n-4} \right] 
\end{align*}
where
\begin{align*}
&\alpha_{1,2} = \frac{1}{2} f''(u) \\
&\Lambda_{n-1,k} = \sum_{i=1}^{n-1} \sum_{j=1}^{2i-1} \alpha_{i,i+j}  \alpha_{n-i,n+k-i-j}, \\
&\Omega_{n-1,j} = (2 n+2 j-1) f''(u) \alpha'_{n-1,n+j-1} + (n+j-1) f'''(u) \alpha_{n-1,n+j-1}.
\end{align*}
\end{theorem}
\proof
Solving the system of linear equations~(\ref{btransport}) by separation of variables one obtains the statement of the theorem.
\endproof
Let us note that coefficients $\alpha$'s depend only on the variable $u$ through the function $f''(u)$ and its higher derivatives. The quasi-Miura transformation for Burgers' equation can be recovered from the formula~(\ref{tburgers}) by choosing $g_{n}(u) = 0$ and eliminating the dependence on function $f''(u)$ and its higher derivatives via the triangular system of the form~(\ref{triang}). It turns out that terms of quasi-Miura transformation at any order are rational functions of $u_x$, $u_{xx}$ etc. with  no explicit dependence on $u$. For instance, up to $O(\epsilon^{4})$, the quasi-Miura transformation for Burgers' equation takes the form
\begin{gather}
\label{qMburgers}
\begin{aligned}
v = &u + \epsilon \frac{u_{xx}}{2 u_{x}} +\epsilon^{2}  \left(\frac{u_{xxx}}{8 u_{x}^{2}} - \frac{u_{xx}^{2}}{6 u_{x}^{3}} \right)_{x} + \epsilon^{3} \left(\frac{u_{5x}}{48 u_{x}^{3}} - \frac{u_{xx} u_{4x}}{6 u_{x}^{4}} - \frac{u_{xxx}^{2}}{8 u_{x}^{4}} + \frac{3 u_{xx}^{2} u_{xxx}}{4 u_{x}^{5}} \right)_{x} + O(\epsilon^{4}).
\end{aligned}
\end{gather}
A direct comparison with the Liu-Zhang approach to the solution of transport equations~\cite{LZ} is readily made by observing that the functions $f''(u)$, $f'''(u)$, $\dots$ play the role of parameters in the quadrature formula~(\ref{Lvnint}) as well as the functions $x_{uu} = f''_{u}$, $x_{uuu} =f'''(u)$, $\dots$ do in Liu-Zhang's framework.

The existence of a transformation relating a viscous scalar conservation law to its inviscid limit allows, at least formally, to obtain solutions of the full hierarchy from solutions obtained via the hodograph formula. Hence, it is natural to look for the viscous counterpart  of this formula. The answer is provided by the following

\begin{theorem}
Given any solution to the Hopf equation
$$u_t=2uu_x,$$
via the hodograph formula~(\ref{Lusol}),
the formal solution of the Burgers equation of the form~\eqref{qMburgers} satisfies the deformed hodograph  equation
\beq\label{hodef}
x+2ut+\omega_f=0
\eeq  
where $\omega_f$ is the deformed 1-form corresponding to $a(u)=1$. 
\end{theorem}

\n
\emph{Proof}. Let us consider the vector field 
\beq\label{}
P\alpha_f=\d_x(x+2ut+f)=1+2u_x\,t+\d_x f.
\eeq
It can be viewed as a linear $t$-dependent combination of the vector fields 
$$\f{\d}{\d u},\,u_x \f{\d}{\d u},\,(\d_x f)\f{\d}{\d u}.$$
Applying to such vector fields the inverse of the quasi-Miura transformation \eqref{qMburgers}
\beq\label{iqMburgers}
u=v+\f{1}{2}\epsilon\frac{v_{xx}}{v_x}+\epsilon^2\left(\f{1}{8}\frac{
v_{(4)}}{v_x^2}-\frac{7}{12}\frac{v_{xxx}v_{xx}}{v_x^3}+\f{1}{2}\frac{v_{xx}^3}{v_x^4}\right)
 +\mathcal{O}(\epsilon^3)
\eeq
and taking into account the transformation law for vector fields
$$X(u)\to\tilde{X}(v)=\left(\f{\d v}{\d u}+\f{\d v}{\d u_x}\d_x+\f{\d v}{\d u_{xx}}\d_x^2+...\right)
X(u)_{|u=u(v,v_x,...)}$$
we can immediately prove that
\begin{itemize}
\item $\f{\d}{\d u}\to\f{\d}{\d v}$ as a consequence of the fact  that $\f{\d v}{\d u}=1$.
\item $u_x\f{\d}{\d u}\to v_x\f{\d}{\d v}$.
\item $(\d_x f)\f{\d}{\d u}\to(\d_x \omega_f)\f{\d}{\d u}$. 
In the general case we only know that such a deformation
  exists up to the order $\epsilon^5$, but in the case of Burgers it is defined for an arbitrary (analytic) function$f$.
 \end{itemize}
Combining the above results we obtain
$$(1+2u_x\,t+\d_x f)\f{\d}{\d u}\to(1+2v_x\,t+\d_x \omega_f)\f{\d}{\d v}$$
This means that, given any hodograph solution $u(x,t)$ the series \eqref{iqMburgers} satisfies
 the equation
\beq
1+2v_x\,t+\d_x \omega_f=0.
\eeq
Integrating with respect to $x$ we obtain
\beq
x+2v\,t+\omega_f=c.
\eeq
where $c$ is a constant. Taking the limit $\epsilon\to 0$ it is immediate to check that the constant $c$ must vanish.

\endproof

Let us now consider the general case where the function $a$ is not constant and suppose, as we have conjectured, that there exists an integrable hierarchy for any choice of the central invariant $a(u)$. 
 As a consequence of the quasi-triviality there should exist also in this case a reducing quasi-Miura transformation. One can easily check that such a transformation depends in general
  on $v$ at any order in the deformation parameter. This immediately implies that the vector
   field $\f{\d}{\d u}$ is no longer invariant and an additional contribution must be taken into account.
If this additional term is a total $x$-derivative, one obtains a correction to the deformed hodograph formula \eqref{hodef}. Indeed if 
 $$\f{\d v}{\d u}=1+\d_x\left(F(u)\right)$$
 then $\f{\d}{\d u}\to[1+\d_x\left(F(u(v))\right)]\f{\d}{\d v}$ and therefore  
$$(1+2u_x\,t+\d_x f)\f{\d}{\d u}\to(1+2v_x\,t+\d_x (\omega_f+F(u(v))))\f{\d}{\d v}.$$
Proceeding exactly as in the above theorem one obtains the deformed hodograph formula
\beq
x+2v\,t+\omega_f+F=0.
\eeq
For example, in the case $a(u)=u$ the reducing transformation is 
\begin{eqnarray*}
v=u+\f{1}{2}\epsilon\left(\frac{uu_{xx}}{u_x}+u_{x}\ln{u_x}\right)
+\f{1}{4}\epsilon^{2}\left(\f{1}{2}
\frac{u^{2}u_{(4)}}{u_x^2}+3\frac{uu_{xxx}}{u_x}-\f{7}{3}\frac{u^2u_{xxx}u_{xx}}{u_x^3}-\f{7}{3}\frac{uu_{xx}^2}{u_x^2}+\right.\\
\left.2\frac{u^2u_{xx}^3}{u_x^4}+\f{1}{2}u_{xx}\left(\ln{u_x}\right)^{2}+2u_{xx}\ln{u_x}+2u_{xx}+\frac{uu_{xxx}\ln{u_x}}{u_x}
-\frac{uu_{xx}^2\ln{u_x}}{u_x^2}\right)+\mathcal{O}(\epsilon^3) 
\end{eqnarray*}
whose inverse is
\begin{eqnarray*}
u=v-\f{1}{2}\epsilon\left(\frac{vv_{xx}}{v_x}+v_x\ln{v_x}\right)-\epsilon^2\left(-\f{1}{8}
\frac{v^{2}v_{(4)}}{v_x^2}-\f{1}{4}\frac{vv_{xxx}}{v_x}+\frac{5}{12}\frac {v^{2}v_{xxx}v_{xx}}{v
_x^3}-\f{1}{12}\frac{vv_{xx}^{2}}{v_x^{2}}+\right.\\
\left.-\f{1}{4}
\frac{v^{2}v_2^3}{v_x^4}-\f{1}{8}v_{xx}
 \left(\ln{v_x}\right)^{2}-\f{1}{2}v_{xx}\ln{v_x} -\f{1}{4}\frac{vv_{xxx}\ln{v_x}}{v_x}+\f{1}{4}
 \frac{vv_{xx}^2\ln{v_x}}{v_x^2}\right)+\mathcal{O}(\epsilon^3).
\end{eqnarray*}
After some computations one gets
\begin{eqnarray*}
&&\f{\d v}{\d u}=1+\d_x\left[\f{1}{2}\epsilon\ln{u_x}+\epsilon^2\left(
\f{1}{4}\f{uu_{xxx}}{u_x^2}+\f{1}{4}\f{u_{xx}\ln{u_x}}{u_x}+\f{1}{2}\f{u_{xx}}{u_x}-\f{1}{3}\f{uu_{xx}^2}{u_x^3}\right)\right]+\mathcal{O}(\epsilon^3)=\\
&&1+\d_x\left[\f{1}{2}\epsilon\ln{v_x}-\epsilon^2\left(
\f{1}{12}\f{vv_{xx}^2}{v_x^3}\right)\right]+\mathcal{O}(\epsilon^3)
\end{eqnarray*}
and therefore the correction to the deformed hodograph formula up to the secon order is given by
$$F(v)=\f{1}{2}\epsilon\ln{v_x}-\epsilon^2\left(
\f{1}{12}\f{vv_{xx}^2}{v_x^3}\right).$$

The asymptotic approach based of the construction of formal solutions via quasi-Miura transformations or equivalently via transport equations is generally expected to provide an accurate
local asymptotic description of solutions to equations of the form~(\ref{claw}) for sufficiently regular initial data and sufficiently small times, i.e.
before the generic solution to the Hopf equation $L u =0$ develops a gradient catastrophe. 
Some results in this direction are already available  for some equations of the form~(\ref{claw}) (see e.g.\cite{MR,MR2}).  As it was pointed out in \cite{MR2}, the fact that the quasi-Miura transformation~(\ref{qM}) does not preserve the initial datum makes difficult to perform a direct comparison between the perturbed and the unpertuberd solution to a specific initial value problem. This problem can be readily fixed using the fact that terms at any order in $\epsilon$ of a quasi-Miura transformation are particular solutions to transport equations~(\ref{transport}) whose general integral is defined up to the kernel of the adjoint operator $L^{*}$. Hence, given a solution to the Hopf equation $Lu=0$ with initial datum $u(x,0) = u_{0}(x)$, the required perturbed solution $v = u + \epsilon v^{1} + \epsilon^{2} v^{2}+ \dots$ that satisfies the same initial datum has to be such that $v^{1}(x,0) =0$,  $v^{2}(x,0) =0$ etc. Such solutions are obtained by choosing the function $h_{n}(u,u_{x}) = g_{n}(u) u_{x}$ in~(\ref{Lvnint}) in such a way that
\[
g_{n}(u_{0}) = - p_{n}(u_{0},u_{0x}) =- p_{n}  \left (u_{0}, - \frac{1}{f'(u_{0})} \right)
\]
 where we have used the formula
\[
u_{0x} = - \left . \frac{1}{2 t + f'(u)}  \right |_{t =0} = - \frac{1}{f'(u_{0})}.
\]

In the KdV case the existence of a
transformation reducing KdV equation to Hopf equation and preserving the initial data was proved in \cite{MR}. The above arguments shows that, in general, the
existence of this transformation relies on the freedom in
the choice of  solutions of transport equations \eqref{Lvn}.

\begin{remark}
Let us recall that the Burgers equation
\[
u_t = 2 u u_{x} + \epsilon u_{xx}
\]
transforms into the heat equation
\begin{equation}
\label{rem_heat}
\varphi_{t} = \epsilon \varphi_{xx}
\end{equation}
via the Cole-Hopf transformation $u = \epsilon \partial_{x} \log \varphi$. Given the initial condition
\[
u(x,0) = F(x)
\]
the corresponding solution to~(\ref{rem_heat}) reads as
\begin{equation}
\label{rem_hsol}
\varphi(x,t) = \frac{1}{\sqrt{4 \pi \epsilon t}} \int_{-\infty}^{\infty} \; e^{-G(x,t,\eta)/2\epsilon} \; d\eta
\end{equation}
where
\[
G(x,t,\eta) = -2 \int_{0}^{\eta} F(s) \; ds + \frac{(x-\eta)^{2}}{2 t}.
\]
For $\epsilon << 1$, the Laplace formula for the asymptotic evaluation of the integral~(\ref{rem_hsol}) gives
\[
\varphi  = \frac{e^{-\frac{G(x,t,\xi)}{2 \epsilon}}}{\sqrt{t G''(x,t,\xi)}} \left(1+ O(\epsilon) \right)
\]
where $\xi= \xi(x,t)$ is a assumed to be the only zero to the equation
\begin{equation}
\label{rem_Gfun}
G'(x,t,\xi) = - 2 F(\xi) - \frac{x-\xi}{t} = 0,
\end{equation}
with the notation $G'(x,t,\xi) = \partial_{\xi} G(x,t,\xi)$.
Up to $O(\epsilon^2)$, we can write
\[
u = \epsilon \partial_{x} \log \varphi \simeq -\frac{1}{2} \partial_{x} G(x,t,\xi) + \epsilon \partial_{x} \log \sqrt{tG''(x,t,\xi)}.
\]
Observing that
\begin{align*}
&\partial_{x}G(x,t,\xi) = - 2 F(\xi) \xi_{x} + \frac{x-\xi}{t} \left(1-\xi_{x} \right) =-2 F(\xi) \\
&t G''(x,t,\xi) =- 2 t F'(\xi) + 1 =\frac{1-\xi_{x}}{\xi_{x}} + 1 = \frac{1}{\xi_{x}}
\end{align*}
we get
\[
u \simeq F(\xi) - \frac{\epsilon}{2} \partial_{x} \log \xi_{x}
\]
Let us now introduce the function
\[
v(x,t) = F(\xi(x,t))
\]
then
\[
\xi = g(v) \qquad \textup{where} \qquad g \equiv F^{-1},
\]
and note that the equation~(\ref{rem_Gfun}) implies that $v$ satisfies the Hopf equation
\[
v_{t} = 2 v v_{x}.
\]
Finally, we obtain the asymptotic formula
\begin{equation}
\label{rem_miura}
u \simeq v  + \frac{\epsilon}{2} \left(\frac{v_{xx}}{v_{x}} + \frac{g''(v)}{g'(v)} v_{x} \right)
\end{equation}
that reproduces at the order $O(\epsilon)$ the quasi-Miura transformation~(\ref{iqMburgers}) up to the solution $h_{0} =\frac{g''(v)}{g'(v)} v_{x}$ of the homogenous transport equation $L^{\ast} h_{0} = 0$ in~(\ref{transport}).
Similarly, higher order corrections in the Laplace formula reproduce higher order terms of the quasi-Miura transformation~(\ref{iqMburgers}). The above calculation shows how the quasi-Miura transformation gives the correct asymptotic formula in the region of the $(x,t)-$plane where the characteristic equation~(\ref{rem_Gfun}) admits one single root. A proof of this fact for a more general class of equations that contain the Burgers equations can be found in~\cite{MR}.

\end{remark}

\section{The Dubrovin-Il'in universality: the viscous analogue of Painlev\'e I2.}

In the Hamiltonian case, it has been conjectured that the behaviour of solutions near the point of gradient catastrophe depends neither on the initial datum (this was proved in \cite{CG} for the KdV equation), nor on the equation (this problem is completely open) and it is governed by a particular solution of Painlev\'e I2 equation. In the non-Hamiltonian case, a similar conjecture has been recently formulated in \cite{DE}. According to this conjecture, the universal behaviour of the solutions near the point of gradient catastrophe is given by the Pearcey integral. This conjecture is supported by a result due Il'in~\cite{I} and it is based on the boundary layer method. 

Based on the results of the previous section, that is combining the deformed hodograph formula together with a suitable double scaling analysis, we derive the viscous analogue of Painlev\'e I2. This extends Dubrovin's work to the case of non-Hamiltonian conservation laws.
 
Let us consider the Hopf equation, that is the conservation law associated with the undeformed 1-form $\omega_{u^{2}} = u^{2}$
\beq
u_{t} = 2 u u_{x},
\eeq
and its solution given in terms of the formula
\beq
\label{hod1}
x  + 2 u t - f(u) = 0
\eeq
where the function $f(u)$ is an arbitrary function that parametrizes the family of commuting flows associated with the unperturbed 1-form $\omega_{f} = f(u)$
\[
u_{\tau}  = f'(u) u_{x}.
\]
The deformation procedure leads to the pair of involutive 1-forms
\begin{align*}
\omega_{u^{2}}^{def} &= u^{2} + \epsilon a  u_{x} + \epsilon^{2} a a' u_{xx} + O(\epsilon^{3}) \\
\omega_{f}^{def} &= f(u) + \frac{\epsilon}{2} a f'' u_{x} + \epsilon^{2} \left [ \left (\frac{1}{2} a a' f'''  +\frac{1}{6} a^{2} f''' \right) u_{xx}  +  \left(\frac{1}{2} a a' f'''+ \frac{1}{8} a^{2} f^{(4)} \right) u_{x}^{2} \right] + O(\epsilon^{3})
\end{align*}
where $a = a(u)$.

Introducing the deformed hodograph equation
\beq
\label{hoddef}
x + 2 u t - \omega_{f}^{def} = 0
\eeq
where $\omega_{f}^{def}$ has been specified for a constant central invariant $a(u) = a_{0}$, we shall perform a multiscale analysis about the
generic point of gradient catastrophe ($x_{0},t_{0},u_{0}$) such that
\beq
\label{critpoint}
x_{0} + 2 u_{0} t_{0} - f(u_{0}) = 0, \qquad 2 t_{0} - f'(u_{0}) = 0, \qquad f''(u_{0}) = 0, \qquad f'''(u_{0}) > 0.
\eeq
Introducing displacement variables ($\ovl{x},\ovl{t},\ovl{u}$) as follows
\beq
x = x_{0} + \lambda^{\alpha} \ovl{x}, \qquad  t = t_{0} + \lambda^{\beta} \ovl{t}, \qquad u = u_{0} + \lambda \ovl{u}
\eeq
where $\lambda = \epsilon^{q}$ is a small parameter, let us now expand in Taylor series the l.h.s of equation~(\ref{hoddef}). Using the conditions~(\ref{critpoint}) one obtains
\begin{gather}
\label{taylor}
\begin{aligned}
\epsilon^{q \sigma} \tilde{x} + 2 \epsilon^{q(\beta+1)} \ovl{u} \ovl{t}  -\frac{\epsilon^{3q} }{6} f'''_{0} \ovl{u}^{3} &- \frac{\epsilon^{1+ q(2-\sigma)}}{2} a_{0} f'''_{0}  \ovl{u}\; \ovl{u}_{\tilde{x}}  - \frac{\epsilon^{2 + q(1-2\sigma)}}{6} a_{0}^{2} f'''_{0} \ovl{u}_{\tilde{x} \tilde{x}}   \\
&= O \left (\epsilon^{4q} \right) + O \left( \epsilon^{1+q (2-\sigma)} \right) + O \left(\epsilon^{2 + q(2-2\sigma)}  \right) 
\end{aligned}
\end{gather}
where the variable $\tilde{x}$ is defined as 
$$
\lambda^{\sigma} \tilde{x} = \lambda^{\alpha} \ovl{x} + 2 u_{0} \lambda^{\beta} \ovl t = x-x_{0} + 2 u_{0} (t-t_{0}),
$$
with the notation  $f_{0} = f(u_{0})$ etc.
The request that all terms into the l.h.s of~(\ref{taylor}) contribute to the same order in $\epsilon$ gives
\[
\sigma = 3 \qquad \beta = 2 \qquad q = \frac{1}{4}.
\]
Hence, we we end up with the expression
\beq
\label{univ1}
\tilde{x} + 2 \ovl{u} \ovl{t}  -\frac{1}{6} f'''_{0} \ovl{u}^{3} - \frac{1}{2} a_{0} f'''_{0}  \ovl{u}\; \ovl{u}_{\tilde{x}}  - \frac{1}{6} a_{0}^{2} f'''_{0} \ovl{u}_{\tilde{x} \tilde{x}}  \\
= O \left (\epsilon^{1/4} \right) .
\eeq
Then, the solution to the deformed hodograph equation in the vicinity of the critical point satisfies the  second order ODE obtained by  taking the limit $\epsilon \to 0$ in~(\ref{univ1}).
The above equation can conveniently be written in non-dimensional form via the following rescaling of the dependent and independent variables
\[
\tilde{x}  = s_{1} X ,\qquad \ovl{t} = s_{2} T, \qquad \ovl{u} = s_{3} U
\]
where
\[
s_{1} = \left(\frac{1}{6} a_{0}^{3} f'''_{0} \right)^{1/4}, \qquad s_{2} = \frac{1}{2} \frac{s_{1}^{2}}{a_{0}} = \left(\frac{1}{24} a_{0} f'''_{0} \right)^{1/2}, \qquad s_{3} =\frac{a_{0}}{s_{1}} = \left(\frac{6 a_{0}}{f'''_{0}} \right)^{1/4}
\]
and $U = U(X,T)$ satisfies the equation
\beq
\label{univ2} 
U_{XX} + 3 U U_{X} + U^{3} - U T = X.
\eeq
In terms of the original variables, the critical behaviour near the point of gradient catastrophe, given a solution to the equation~(\ref{univ2}), is
\beq
\label{criticalb}
u = u_{0} + \left(\frac{6 a_{0}}{f'''_{0}} \right)^{1/4} \epsilon^{1/4} U \left [ \left(\frac{6}{a_{0}^{3} f'''_{0}} \right)^{1/4} \frac{x-x_{0} + 2 u_{0} (t-t_{0})}{\epsilon^{3/4}}, \left ( \frac{24}{a_{0} f'''_{0}} \right)^{1/2} \frac{t-t_{0}}{\epsilon^{1/2}} \right ].
\eeq
We also observe that the equation~(\ref{univ2}) can be transformed via a Cole-Hopf transformation of the form $U = \partial_{X} \log w$ into the linear ODE
\beq
\label{univ2lin}
w_{XXX} - T w_{X} = X w.
\eeq
Equation \eqref{univ2lin} has the following general solution 
$$w(X, T)=c_1\;
{\mbox{$_0$F$_2$}([1/2,3/4],\,{\frac {1}{64}}\, \left( T+X \right) ^{4})}
+c_2\; {\mbox{$_0$F$_2$}([3/4,5/4],\,{\frac {1}{64}}\, \left( T+X \right) ^{4})}
 \left( T+X \right) 
$$
$$+c_3\;{\mbox{$_0$F$_2$}([5/4,3/2],\,{\frac {1}{64}}\, \left( T+X \right) ^{4})}\left( T+X \right) ^{2},$$
where $c_1, c_2, c_3$ are arbitrary integration constants and $\mbox{$_0$F$_2$}$ is the generalized hypergeometric function defined by 
\beq
\label{hyper}
\mbox{$_0$F$_2$}([\alpha,\beta],\,z):=\sum_{n=0}^{+\infty}\frac{1}{(\alpha)_n (\beta)_n}\frac{z^n}{n!},
\eeq
where $$(\alpha)_n:=(\alpha)(\alpha+1)(\alpha+2)\dots(\alpha+n-1), \quad (\alpha)_0:=1$$
and analogously for $(\beta)_n.$ The function $\mbox{$_0$F$_2$}([\alpha,\beta],\,z)$ defines an entire function whenever $\alpha>0$ and $\beta>0$ (it is clear from the expansion (\ref{hyper}) that is dominated by the series of the exponential function up to a constant).

According to the conjecture formulated in \cite{DE} that generalizes a result by A. Il'in \cite{I},  the critical behaviour~(\ref{criticalb}) is provided by the particular solution to the equation~(\ref{univ2})
\beq
U =  \partial_{X} \log P(X,T) =  \frac{P_{X}(X,T)}{P(X,T)}
\eeq
where $P(X,T)$ is the Pearcey integral
\[
P(X,T)  = \int_{-\infty}^{\infty} e^{- (4 z^{4} - 2 T z^{2} + 2 X z)} \; dz.
\]
By a direct calculation using the identity
\[
\int_{-\infty}^{\infty} \left(-16 z^{3} + 4 T z - 2 X \right) \; e^{- \left (4 z^{4} -2 T z^{2} + 2 X z \right)} \; dz = 0
\]
one can directly verify that the function $P(X,T)$ satisfies the linear ODE~(\ref{univ2lin}).

\subsubsection*{Aknowledgements} The authors wish to thank Roberto Camassa, Boris Dubrovin and Gregorio Falqui for useful discussions and references.
 The research of P.L. is partially supported by the Italian MIUR Research Project \emph{Teorie geometriche e analitiche dei sistemi Hamiltoniani in dimensioni finite e infinite} and by  GNFM Programme \emph{Progetto Giovani 2012}. The research of A.M. is partially supported  by the ERC grant \emph{FroM-PDE}. A.M. is also grateful to B. Dubrovin and T. Grava for the kind hospitality at SISSA in Trieste.

\end{document}